\documentclass[wrr]{agutex}

 \usepackage{graphicx}
\usepackage{amsmath,amssymb}
\usepackage{bm}
\usepackage{upgreek}

\usepackage{color}

\authorrunninghead{SALESKY ET AL.}

\titlerunninghead{TRANSPORT AND DEPOSITION OF HEAVY PARTICLES}

\authoraddr{Corresponding author: S. T. Salesky,
School of Meteorology, University of
Oklahoma, 120 David L.~Boren Blvd. \#5900, Norman, OK 72073.
(salesky@ou.edu)}

\begin{document}

\newcommand{\sts}[1]{\textcolor{magenta}{#1}}

\title{The transport and deposition of heavy particles in complex terrain:
  insights from an Eulerian model for large eddy simulation}


\authors{S. T. Salesky\altaffilmark{1}, M. G. Giometto\altaffilmark{2},
    M. Chamecki\altaffilmark{3}, M. Lehning\altaffilmark{2,4}, and 
    M. B. Parlange\altaffilmark{6}}

\altaffiltext{1}{School of Meteorology, University of Oklahoma, 
    Norman, OK, USA}

\altaffiltext{2}{Department of Civil Engineering and Engineering Mechanics,
    Columbia University, New York, NY, USA}

\altaffiltext{3}{Department of Atmospheric and Oceanic Sciences, University of California Los Angeles,
    Los Angeles, CA, USA}

\altaffiltext{4}{School of Architecture, Civil and Environmental Engineering, 
    Swiss Federal Institute of Technology, Lausanne, Switzerland}

\altaffiltext{5}{WSL Institute for Snow and Avalanche Research SLF, Davos, Switzerland}

\altaffiltext{6}{Department of Civil Engineering, Monash University, Clayton,
    Australia}

\begin{abstract}

The transport and deposition of heavy particles over complex surface topography
by turbulent fluid flow is an important problem in a number of disciplines,
including sediment and snow transport, ecology and plant pathology, aeolian
processes, and geomorphology. This article presents a framework to simulate the
transport and deposition of heavy particles over complex surfaces using the
large eddy simulation (LES) technique. An immersed boundary LES code is coupled
with an Eulerian particle code that solves the advection-diffusion equation for
the resolved particle concentration field. The mass conservation equation for
the particle phase is discretized in a finite volume framework using a
Cartesian cut cell method that reshapes finite volume cells intersected by the
immersed boundary surface and conserves mass accurately. The proposed numerical
model is compared with data from wind tunnel experiments of heavy particle
deposition over topography and is found to have good agreement with observed
deposition patterns. An LES case study of snow deposition over idealized topography
leads to several new insights. Particle inertia leads to relative velocities
between the particles and fluid in regions of mean flow acceleration, thereby
enhancing deposition on the windward side of obstacles and suppressing
deposition on the leeward side. In addition, it is found that the mean
components of particle inertia are a factor of 6 or more larger than the
turbulent components, indicating that the enhancement/suppression of deposition
by topography can be modeled in terms of mean flow quantities.

\end{abstract}

\begin{article}

\section{Introduction}\label{sec:intro}

The transport, erosion, and deposition of heavy particles over complex surface
topography due to turbulent flow is an important process for a diverse set of
research problems, including alpine hydrology and ecology; avalanche
prediction; dune formation and evolution; the dispersal of pollen, seeds, and spores;
the emission of mineral dust aerosols;
stream bank erosion; and scour around hydraulic structures.

Snow deposition, erosion, and transport over complex topography are important
processes in the context of alpine and polar hydrology, ecology, and avalanche
prediction. A number of studies have demonstrated that the spatial distribution
of snowpack depth and snow water equivalent are closely linked to topography
\citep[e.g.][]{elder1991snow, bloschl1992analysis, liston1998snow,
  luce1998influence, balk1999distribution, lehning2008inhomogeneous}.  Spatial
inhomogeneities in snowpack depth have a strong influence on the timing of
surface runoff through snowpack melt; \citet{luce1998influence} found that
errors in estimates of snowpack depth led to basin-wide errors in runoff and
evapotranspiration in distributed hydrological models. Runoff from snow melt is
a major source of nutrients for alpine ecosystems
\citep{bowman1992inputs,walker1993long}, and snowpack depth can influence the
spatial variability of available moisture, which is a determining factor for
the spatial distribution of vegetation cover \citep{evans1989spatial}.  In
addition, the inhomogeneous transport and deposition of snow in complex terrain
can be a contributing factor for avalanche formation
\citep[e.g.][]{perla1976avalanche,schweizer2003snow}.  Snow cornices
\citep{kobayashi1988formation,vogel2012cornice} frequently form on the leeward
side of ridges; cornice failure often triggers avalanche formation.
Furthermore, snow transport is a significant process influencing the mass
balance of ice sheets
\citep[e.g.][]{eisen2008ground,scarchilli2010extraordinary,
  das2013influence,das2015extreme} and continental glaciers
\citep{dery2010blowing}.

In the context of sediment transport, saltating sand particles can be a source
of mineral dust aerosols \citep{shao1993effect}, which have important
implications for climate, ecology, and hydrology
\citep[e.g.][]{shao2011dust,kok2012physics}. Radiative feedbacks due to dust
remain a significant source of climate uncertainty \citep{sokolik2001introduction}.
Aeolian dust transport to the oceans is a significant source of nutrients (e.g. iron)
that affect ocean ecology and biogeochemistry \citep{jickells2005global}. In
addition, mineral dust aerosols have complex feedbacks on the hydrological
cycle, leading to surface radiative forcings that reduce global precipitation,
but increase precipitation in arid regions \citep{miller2004surface}.

Interactions between a turbulent flow and mobile sediment bed can lead to the
formation of a variety of bedforms including ripples and dunes
\citep[e.g.][]{bagnold1941physics,engelund1982sediment,charru2013sand}; recent
studies have demonstrated how these bedforms can propagate as sand waves
\citep[e.g.][]{venditti2005bed,khosronejad2014numerical}. The formation and
migration of these bedforms in rivers can affect stream ecology
\citep{macvicar2006two} and can cause riverbank erosion during floods
\citep{best2005fluid,macvicar2006two}.  In addition, the presence of bedforms
in rivers can also enhance the exchange of solutes and particles in the
hyporheic zone \citep{packman2001hyporheic, packman2004hyporheic} which impact
contaminant transport and stream ecology.  Furthermore, coherent
structures in a turbulent flow can lead to scour around hydraulic structures such as bridge piers
\citep{khosronejad2012experimental} and can lead to bridge failure
\citep{briaud1999sricos}. 

The transport of biogenic particles such as pollen \citep{di1991factors},
fungal spores \citep{aylor1990role,brown2002aerial}, and seeds
\citep{nathan2002mechanisms} is also a relevant question for ecology and plant
pathology. The wind-borne dispersion of seeds is important for ecological
issues such as gene flow, plant colonization, and the spread of invasive
species \citep[e.g.][]{cain2000long,nathan2000spatial,nathan2002mechanisms}.
Pollen dispersion from field crops has become a topic of interest in recent
years, in order to quantify the probability of gene flow from genetically
modified (GM) to non-GM wind-pollinated crops such as maize (\textit{Zea mays})
\citep{aylor2002settling,aylor2003aerobiological}. The turbulent dispersion of
biogenic gases, such as pheromones, is known to be significant for the
lifecycle of insect pests, such as the gypsy moth (\textit{Lymantria
  dispar}) \citep{aylor1976turbulent}.  In addition, many diseases
affecting field crops are carried by fungal spores \citep{brown2002aerial},
which may be transported long distances after being entrained from plant leaves
or stalks \citep{aylor1975ventilation}. 

While some regional-scale studies have demonstrated that local flow patterns
caused by topography can enhance the long-distance dispersion of heavy
particles such as pollen \citep[e.g.][]{helbig2004numerical}, and recent large
eddy simulation (LES) studies of particle dispersion within and above plant
canopies have considered complexities such as finite size area sources
\citep{chamecki2012analytical}, unstable stratification
\citep{pan2013dispersion}, plant reconfiguration \citep{pan2014large}, and edge
effects \citep{pan2015dispersion}, many question regarding the effects of
topography on the dispersal of biogenic particles remain to be explored.

A variety of approaches have been employed in previous numerical studies of the
transport of heavy particles over topography.  Many previous studies
\citep[e.g.][]{demuren1986calculation, wu20003d, zedler2001large,
  gauer2001numerical, nagata2005three, roulund2005numerical,
  ortiz2006numerical, ortiz2009coupling} have used boundary-fitted grids. When
this approach is used for simulations where the surface is allowed to evolve,
remeshing will be required every time the surface deforms due to erosion or
deposition.  A number of numerical issues are associated with remeshing,
including decreased numerical accuracy as the simulation progresses
\citep{khosronejad2011curvilinear}. Furthermore, grid generation can be a
challenge in complex domains \citep{mittal2005immersed,
  khosronejad2011curvilinear}.

Other studies of particle transport in turbulent flows have employed a
Lagrangian approach where the governing equations of particle motion are solved
directly for an ensemble of particles \citep[e.g.][]{shao1999numerical,
  vinkovic2006large, dupont2013modeling, zwaaftink2014modelling,
  finn2016particle}.  While this approach is advantageous for studying
near-surface processes in detail (e.g.  particle entrainment, saltation, and
rebound), Lagrangian methods currently are too expensive to use for studies of
bedform evolution \citep{sotiropoulos2016sand} due to the large number of
particles ($\sim$billions) required to converge statistics; the saltation
layer in drifting snow or sand can carry on the order of $1.5 \times 10^6$
particles m$^{-2}$ \citep{gauer2001numerical}.

Although immersed boundary LES of sediment transport has been conducted
recently in an Eulerian framework \citep{khosronejad2014numerical}, most
studies of snow transport to date have been conducted following the
Reynolds-averaged Navier Stokes (RANS) approach, where all scales of turbulence
are parameterized \citep[e.g.][]{gauer2001numerical, lehning2008inhomogeneous,
  schneiderbauer2011atmospheric} or in LES using Lagrangian particles
\citep{zwaaftink2014modelling}. 

In this article we present a new approach for modeling the transport and
deposition of heavy particles over surface topography in LES---applicable to a
diverse set of  problems including snow and sediment transport and the
dispersion of heavy particles and passive scalars in complex terrain and urban
environments.  An immersed boundary version of an existing large eddy
simulation code \citep{albertson1999surface,kumar2006large} for momentum is
coupled with a Eulerian particle code \citep[][]{chamecki2009large} that solves
the advection-diffusion equation for heavy particles that includes
gravitational settling and inertia.  The particle mass conservation equation is
discretized in a finite volume framework. In order to ensure mass conservation,
we employ a Cartesian cut cell method \citep[e.g.][]{udaykumar1996elafint,
  ye1999accurate, udaykumar2001sharp, ingram2003developments,
  mittal2005immersed} where the finite volumes intersected by the fluid-solid
interface are reshaped, i.e. the face area and volume fractions of a cut cell
that lie in the fluid are accounted for explicitly in the discretized version
of the mass conservation equation. We also modify the conservative
interpolation scheme developed by \citet{chamecki2008hybrid} to ensure that the
interpolated velocity on the faces of the cut cells remains divergence-free. In
addition, the wall models for particle deposition and erosion are modified to
account for the possibility of a sloping surface.

This article is organized as follows. A description of the large eddy
simulation model is presented in Sec.~\ref{sec:model}, including the new
developments required to accurately simulate particle transport and deposition
over surface topography.  In Sec.~\ref{sec:validation}, a validation case is
presented where LES results are compared with wind tunnel experiments of
particle deposition over topography.  In Sec.~\ref{sec:snow_deposition}, we
present a case study of snow deposition over idealized surface topography, and
use the LES to investigate how topography influences deposition patterns.  A
discussion of our results and the main conclusions will be presented in
Sec.~\ref{sec:conclusions}.

\section{Model description}\label{sec:model}

\subsection{Large Eddy Simulation Code}

The large eddy simulation code used in this study (discussed in detail by
\citet{albertson1999surface} and \citet{kumar2006large}) solves the
three-dimensional filtered momentum equation written in rotational form.  The
governing equations are discretized using a pseudospectral collocation approach
for horizontal derivatives, and second-order centered finite differences in the
vertical, with the fully-explicit second-order Adams Bashforth method used for
time integration. Nonlinear terms are fully dealiased using the 3/2 rule
\citep[e.g.][]{canuto2012spectral}. The resulting system of algebraic equations
is solved using a fractional step method \citep{chorin1968numerical}.  In the
simulations considered herein, we will compare the performance of several
subgrid models in the context of particle transport and deposition, including
the static Smagorinsky model \citep{smagorinsky1963general}, the plane-averaged
dynamic model \citep[][]{lilly1992proposed}, and the Lagrangian
scale-dependence dynamic (LASD) model \citep{bouzeid2005scale}, which applies
the dynamic procedure \citep{lilly1992proposed} by averaging over Lagrangian
trajectories of fluid parcels \citep{meneveau1996lagrangian} to determine a
value of the Smagorinsky coefficient while relaxing the scale invariance
assumption of the model coefficient $c_s$, which is especially important in
the near-wall region.  Periodic boundary conditions are used in the horizontal
directions; at the domain top, stress-free and zero vertical velocity
conditions are imposed.  The wall model (used over flat, horizontal homogeneous
surfaces with no immersed boundary) is based on imposing Monin-Obukhov
similarity in a local sense \citep{kumar2006large} with a test filter applied
at scale $2\Delta$ to better reproduce the mean surface stress
\citep{bouzeid2005scale}. In the present work we perform neutrally-stratified
simulations of a turbulent half-channel, forcing with a constant pressure
gradient force in the streamwise direction.

\subsection{Immersed Boundary Method}\label{sec:ibm_method}

\begin{figure}[htbp]
  \begin{center}
    \includegraphics[scale=1.0]{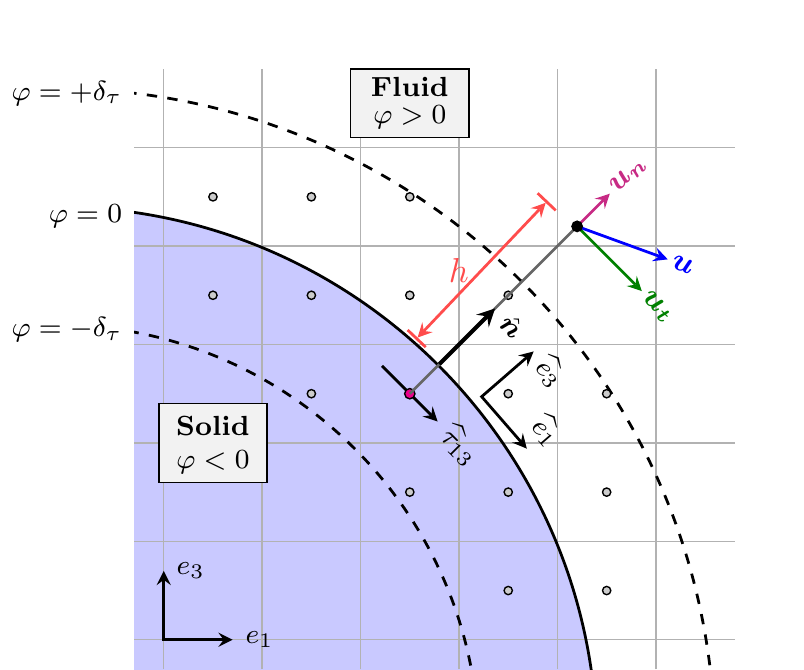}
  \end{center}
  \caption{A schematic diagram illustrating the immersed boundary method for
    momentum. The wall model is applied in a band on nodes in the range
    $-\delta_\tau \leq \varphi \leq +\delta_\tau$, where $\varphi$ is the level
    set function, $\bm{\widehat{n}}$ is the surface normal vector, $\bm{e_i}$
    and $\bm{\widehat{e}_i}$ represent the basis vectors in the Cartesian and
    local coordinate systems respectively, $\bm{u}$ is the velocity vector
    calculated by trilinear interpolation at distance $h$ from the wall,
    $\bm{u_n} = (\bm{u} \cdot \widehat{\bm{n}})\,\widehat{\bm{n}}$ is the normal
    velocity component, $\bm{u_t} =
    \bm{u}-(\bm{u}\cdot\bm{\widehat{n}})\,\bm{\widehat{n}}$ is the tangential
    velocity component, and $\widehat{\tau}_{13}$ is the surface shear stress
    in the local coordinate system.}
  \label{fig:ibm_momentum}
\end{figure}

Surface topography is represented in the LES using an immersed boundary method
(IBM) following the discrete direct forcing approach
\citep{mohd1997combined,mittal2005immersed}.  The use of an immersed boundary
method, as opposed to other methods of representing surface topography (such as
terrain-following coordinates or unstructured grids), has the advantages of
easy and computationally inexpensive implementation, since the underlying
discretization is still done on a Cartesian grid.  Our implementation of the
IBM is similar to that employed by \citet{chester2007modeling}, and will be
summarized below.

The interface between the fluid and the solid is represented in the code using
a level set ($\varphi$), a signed distance function. The level set is zero at
the solid-fluid interface ($\varphi = 0$), negative inside the solid
($\varphi < 0$), and positive in the fluid ($\varphi > 0$). The 
outward-facing normal vector on the surface is related to the level set
function via
\begin{equation}\label{eq:n_hat} 
  \bm{\widehat{n}} = \frac{\nabla \varphi}{|\nabla \varphi|}  \text{.}
\end{equation}

Because the LES is not a wall-resolving simulation, a wall model must be used
in the region near the immersed boundary interface in order to recover the
correct velocity profile.  The wall model for the surface shear stress  is
implemented in the band $-\delta_\tau \leq \varphi \leq +\delta_\tau$, where
$\delta_\tau = 1.1\, \Delta z$.  The wall model in this band is imposed
according to the following steps (a schematic diagram can be found in
Fig.~\ref{fig:ibm_momentum}): 

\textit{Step 1} For each grid node in the band $|\varphi| \leq \delta_\tau$,
the velocity vector $\bm{u}$ is calculated at a point $h = 1.5 \, \Delta z $
away from the wall in the direction of the normal vector $\bm{\widehat{n}}$ using trilinear
interpolation.

\textit{Step 2} A local
(wall-relative) coordinate system is defined using $\bm{\widehat{n}}$ and the
tangential velocity component $\bm{u_t} = \bm{u} - (\bm{u} \cdot
\bm{\widehat{n}}) \, \bm{\widehat{n}}$, via:
\begin{equation}
( \bm{\widehat{e}_1} ,\; \bm{\widehat{e}_2} ,\; \bm{\widehat{e}_3} ) = 
\left( \frac{\bm{u_t}}{|\bm{u_t}|} ,\; \bm{\widehat{n}} \times \bm{\widehat{e}_1} ,\; \bm{\widehat{n}} \right) \text{.}
\end{equation}

\textit{Step 3} The SGS shear stress in this local coordinate system is
calculated as:
\begin{equation}\label{eq:local_shear_stress}
  \widehat{\tau}_{13} = - \rho \left[ \frac{\kappa |\bm{u_t}| }{\ln (h/z_0)}  \right]^2\text{,}
\end{equation}
where $\kappa = 0.4$ is the von K\'{a}rman constant, $z_0$ is the roughness
length, and $h = 1.5 \, \Delta z$ is the distance from the wall.

\textit{Step 4} The SGS shear stress is transformed back to the Cartesian
coordinate system of the LES grid via:
\begin{equation}
\tau_{ij} = a_{in} \widehat{\tau}_{nm} a_{mj}
\end{equation}
where $a_{in}$ and $a_{jm}$ are the direction cosines, e.g. $a_{in} = \bm{e_i}
\cdot \bm{\widehat{e}_n}$, and $(\bm{e_1}, \bm{e_2}, \bm{e_3})$ are the basis
vectors in the Cartesian frame of reference.

At each timestep, the velocity field inside the immersed boundary ($\varphi <
0$) is set to zero, then polynomial smoothing is applied inside the IBM surface
to reduce the Gibbs oscillations that arise due to the spectral differentiation in
horizontal directions. This is done prior to the pressure solver, which enforces 
incompressibility ($\nabla \cdot \bm{u} = 0$) of the velocity field.

\subsection{Finite-Volume Particle Code}

In order to avoid the large computational overhead associated with including a
large number of Lagrangian particles in the LES required to converge
statistics, an Eulerian framework is adopted here, an approach that has been
used successfully to model diverse problems including the dispersion of heavy
particles in the atmospheric boundary layer \citep{dupont2006eulerian,
  chamecki2009large, chamecki2011particle, pan2013dispersion, freire2016flux},
sediment transport \citep{zedler2001large, chou2008modeling,
  khosronejad2014numerical}, gravity currents \citep{necker2002high}, snow
transport \citep{lehning2008inhomogeneous}, and oil plumes in ocean turbulence
\citep{yang2015oil, chen2016plumes}.  The evolution of the particle
concentration field is represented using an advection-diffusion
equation, given as: 
\begin{equation}\label{eq:pcon}
  \frac{\partial \widetilde{C}}{\partial t} +\nabla \cdot (\bm{\widetilde{v}_p}\widetilde{C}) = - \nabla \cdot \bm{\pi}^C + Q_{src} \text{.}
\end{equation}
Here $\widetilde{C}$ is the resolved particle concentration field, $\widetilde{\bm{v}}_p$
is the particle advection velocity, $\bm{\widetilde{u}}$ is the resolved
velocity field, $Q_{src}$ is a source term (e.g. for point or area source
releases), and $\bm{\pi^C}$ is the SGS particle concentration flux, 
defined as
\begin{equation}
  \bm{\pi}^C = \widetilde{\bm{u} C} - \widetilde{\bm{u}}\widetilde{C} \text{,}
\end{equation}
and modeled using a flux-gradient model:
\begin{equation}
  \bm{\pi}^C_{model} =  - \frac{\nu_{sgs}}{{Sc}_{sgs}} \nabla \widetilde{C} \text{,}
\end{equation}
where $\nu_{sgs} = (c_s \Delta)^2 |\widetilde{\bm{S}}|$ is the SGS eddy
viscosity, $\Delta$ is the LES filter width,
$\widetilde{S}_{ij} =
\frac{1}{2}\left(\frac{\partial \widetilde{u}_i}{\partial x_j} + \frac{\partial
    \widetilde{u}_j}{\partial x_i} \right)$ is the filtered strain rate tensor,
$|\widetilde{\bm{S}}|=(2\widetilde{S}_{ij}\widetilde{S}_{ij})^{1/2}$ is its magnitude, 
$c_s$ is the Smagorinsky coefficient, and ${Sc}_{sgs}$ is the SGS Schmidt
number, which for our present simulations is assumed to be constant.  The
particle advection velocity is defined as 
\begin{equation}\label{eq:advection_vel}
  \bm{\widetilde{v}_p} = \widetilde{\bm{u}} + \tau_p \bm{g} - \tau_p \widetilde{\bm{a}} \text{,}
\end{equation}
where the terms on the right hand side represent contributions from the fluid
velocity, gravitational settling, and particle inertia respectively. Here
$\tau_p$ is a characteristic particle timescale, $\bm{g}$ is the gravity
vector, and $\widetilde{\bm{a}}$ is the particle acceleration vector.  The
gravitational settling term is expressed  as $\tau_p \bm{g} = -w_s \bm{e_3}$,
where $w_s$ is a mean settling velocity for the particles, which in the present
study are assumed to be monodisperse. Thus, the particle timescale is given as
$\tau_p = w_s/g$.

The inertia term is modeled by replacing the particle acceleration by the fluid
acceleration
\citep[e.g.][]{shotorban2007eulerian,balachandar2010turbulent}, i.e.
\begin{equation}\label{eq:fluid_acc}
  \widetilde{\bm{a}} = \frac{D \widetilde{\bm{u}}}{Dt} = - \widehat{\nabla
    \widetilde{p^*}} + \frac{1}{2}  \nabla \left(\widetilde{\bm{u}}\cdot
    \widetilde{\bm{u}} \right), 
\end{equation}
where $\widetilde{p^*}$ is the modified pressure that enforces
incompressibility.  Here $(\,\widehat{\cdot}\,)$ denotes the test filtering
operation at scale $2\Delta$; we find that in simulations of inertial particles
over an immersed boundary, test filtering of the fluctuating pressure gradient
term reduces nonphysical oscillations that occur in particle deposition (due to
the Gibbs oscillations that appear in the modified pressure and its
derivatives).  Note that the contribution from the divergence of the SGS stress
tensor is neglected in \eqref{eq:fluid_acc} due to the fact that only a small
amount of energy is contained in the SGS scales, and because $\tau_p$ is small
in the equilibrium assumption \citep[e.g.][]{shotorban2007eulerian,
  yang2016large}.

Because \citet{chamecki2009large} were interested in the dispersion of heavy
particles from point or area sources (which can lead to strong spatial
gradients), they discretized \eqref{eq:pcon} in a finite volume framework
rather than employing the pseudospectral approach often used in LES.  The
discrete version of \eqref{eq:pcon} on a Cartesian grid over a flat surface
(i.e. with no immersed boundary), written here for one Euler time step, is
given as:
\begin{multline}\label{eq:pcon_discrete}
  \frac{\widetilde{C}^{n+1}_{i,j,k} - \widetilde{C}^n_{i,j,k}}{\Delta t} =   \\
  - \frac{1}{\Delta x} \left[ F_{i+1/2,j,k} - F_{i-1/2,j,k} \right]  
  - \frac{1}{\Delta y} \left[ F_{i,j+1/2,k} - F_{i,j-1/2,k} \right]  \\
  - \frac{1}{\Delta z} \left[ F_{i,j,k+1/2} - F_{i,j,k-1/2} \right] 
  + \frac{Q_{src}}{V_{cell}}\text{,}
\end{multline}
where $\widetilde{C}^{n}_{i,j,k}$ is the concentration in cell $(i,j,k)$ at
time $t = n \,\Delta t$, $F_{i,j,k}$ denotes the sum of the advective and SGS
diffusive fluxes on a face of the control volume, and $V_{cell} = \Delta x \,
\Delta y \, \Delta z$ is the volume of a cell.  The face-averaged fluxes are
calculated as (e.g. for the east $x$-face of a control volume):
\begin{multline}
F_{i+1/2,j,k} = \left[ U_{i+1/2,j,k} \, C_{i+1/2,j,k} \right] -   \\
\left[K_{i+1/2,j,k} \, \frac{1}{\Delta \, x} \left(C_{i+1,j,k} - C_{i,j,k}   \right)  \right]
\end{multline}
where the first term is the advective flux, the second term is the SGS
diffusive flux, and $K_{i,j,k} = \nu^{sgs}_{i,j,k}/{Sc}_{sgs}$ is the SGS
diffusivity for particles.

The interpolation of particle concentration to the faces of the control volumes
is done using SMART, a bounded third-order upwind scheme proposed by
\citet{gaskell1988curvature}, which prevents nonphysical negative
concentrations. The velocity interpolation is done using the conservative
interpolation scheme proposed by \citet{chamecki2008hybrid}, which ensures that
the interpolated velocity field on the finite volume faces remains
divergence-free.  A discussion of how the finite volume spatial discretization
is modified for use with the immersed boundary method can be found in
Sec.~\ref{sec:cut_cell}.

The lower boundary condition for particle concentration is derived based on the
assumption of equilibrium between gravitational settling and turbulent
diffusion \citep{chamberlain1967transport,kind1992one}, which was later
modified by \citet{freire2016flux} to account for thermal
stratification. In the present study, we consider only neutral
stratification. The total particle concentration flux at the wall can be
separated into a deposition term and a source term \citep{chamecki2009large}:
\begin{equation}
  \Phi_{sgs}(x,y) = \Phi^{src}_{sgs}(x,y) + \Phi^{dep}_{sgs}(x,y) \text{,}
\end{equation}
where
\begin{equation}\label{eq:pcon_src}
  \Phi^{src}_{sgs}(x,y) = w_s C_r \left( \frac{z_1}{z_r}
  \right)^{-\gamma} \left[ 1 - \left( \frac{z_1}{z_r}
    \right)^{-\gamma}  \right]^{-1} 
\end{equation}
and
\begin{equation}\label{eq:pcon_dep}
  \Phi^{dep}_{sgs}(x,y) = - w_s \widetilde{C}(x,y,z_1)  \left[ 1 - \left(
      \frac{z_1}{z_r} \right)^{-\gamma} \right]^{-1} \text{.}
\end{equation}
In \eqref{eq:pcon_src}--\eqref{eq:pcon_dep}, $z_1 = \Delta z / 2$ is the height
of the first node where $\widetilde{C}$ is stored 
(in the LES we employ a staggered grid arrangement, where $\widetilde{C}$ is
collocated with the $u$ and $v$ velocity components and pressure, whereas $w$
is stored $\Delta z / 2$ above and below the $uvp$ nodes), 
and $z_r$ is a height where a reference concentration
$C_r$ is imposed. Here we take $z_r = z_{0,c}$, i.e. we use the roughness
height for particle concentration as the reference height. The quantity
$\gamma$, given as 
\begin{equation} 
  \gamma = \frac{{Sc}_T w_s}{\kappa u_*}
\end{equation} 
is a dimensionless parameter (known as the Rouse number) that represents the
relative importance of turbulent diffusion and gravitational settling, where
${Sc}_T$ is the turbulent Schmidt number, which is not necessarily 
equal to the SGS Schmidt number ${Sc}_{sgs}$. Here we adopt a value of ${Sc}_T
= 0.95$, which in the limit $w_s \rightarrow 0$, recovers the commonly accepted
form \citep{hogstrom1988non} of the  Monin-Obukhov similarity function for
scalars \citep{chamecki2009large}.

\subsection{Cartesian Cut Cell Method for Particle Concentration}\label{sec:cut_cell}

In order to ensure explicit conservation of mass in the particle phase,
we use a Cartesian cut cell method to discretize the particle concentration
equation in the finite volume framework when an immersed boundary is present.
In Cartesian cut cell methods \citep[e.g.][]{udaykumar1996elafint,
  ye1999accurate, udaykumar2001sharp, ingram2003developments,
  mittal2005immersed}, finite volume cells intersected by the immersed boundary
surface (i.e. the zero level set) are reshaped to ensure that the integral form of
the conservation equation is satisfied explicitly in these irregularly shaped
(non-Cartesian) cells. This is done by introducing the volume fraction of each
cell and the area fraction of each face in the fluid into the discretized version
of the particle mass conservation equation.

\subsubsection{Discretization in Cut Cell Framework}

\begin{figure}[htbp]
  \begin{center}
    \includegraphics[scale=1.0]{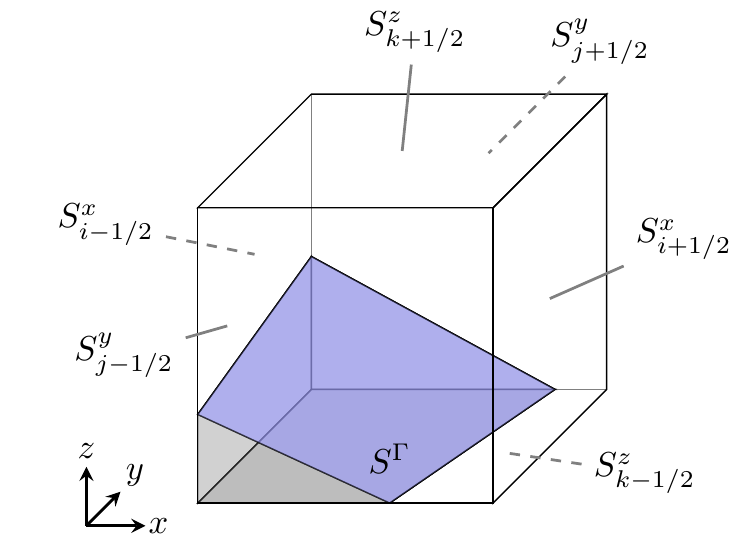}
  \end{center}
  \caption{A schematic diagram of the regular face ($S_x$, $S_y$, and $S_z$) and cut face ($S_\Gamma$) areas in a cut cell.}
  \label{fig:cut_cell}
\end{figure}

In the Cartesian cut cell framework, the discrete version of the particle mass
conservation equation \eqref{eq:pcon_discrete}, again written for one Euler
step, becomes:
\begin{multline}\label{eq:pcon_cut_cell}
  \frac{\widetilde{C}^{n+1}_{i,j,k} - \widetilde{C}^n_{i,j,k}}{\Delta t} =   \\
  - \frac{1}{\alpha_{i,j,k} \Delta x} \left[A^{x}_{i+1/2,j,k} F_{i+1/2,j,k} - A^{x}_{i-1/2,j,k} F_{i-1/2,j,k} \right]  \\
  - \frac{1}{\alpha_{i,j,k} \Delta y} \left[ A^{y}_{i,j+1/2,k} F_{i,j+1/2,k} - A^y_{i,j-1/2,k} F_{i,j-1/2,k} \right]  \\
  - \frac{1}{\alpha_{i,j,k} \Delta z} \left[ A^{z}_{i,j,k+1/2} F_{i,j,k+1/2} - A^z_{i,j,k-1/2} F_{i,j,k-1/2} \right]   \\
  + \frac{Q_{src}}{\alpha_{i,j,k} V_{cell}} 
  + \frac{S^{\Gamma}_{i,j,k}}{\alpha_{i,j,k} V_{cell}} (\widehat{\Phi}^{dep}_{sgs} + \widehat{\Phi}^{src}_{sgs} ) \text{.}
\end{multline}
In \eqref{eq:pcon_cut_cell}, $\alpha_{i,j,k}$ is the volume fraction of a cut
cell in the fluid, i.e. $\alpha = 1$ for a regular cell fully in the fluid, $0
< \alpha < 1$ for a cut cell, and $\alpha = 0$ for a cell fully in the solid.
(Note that, in practice, the right-hand side of \eqref{eq:pcon_cut_cell} is set
to zero at each timestep in the code in cells where $\alpha = 0$ to avoid
division by zero).  The fluxes on the finite volume faces are modified by the
fraction of the face in the fluid, e.g. $A^x_{i+1/2,j,k}$ for the east
$x$-face, where $0 \leq A^x , A^y, A^z \leq 1$. The relationship between the
face areas ($S^x$, $S^y$, and $S^z$) and the face area fractions is
\begin{equation}
\begin{aligned}
  S^x_{i,j,k} = \Delta y \, \Delta z \, A^x_{i,j,k} \\
  S^y_{i,j,k} = \Delta x \, \Delta z \, A^y_{i,j,k} \\
  S^z_{i,j,k} = \Delta x \, \Delta y \, A^z_{i,j,k}  \text{.}
\end{aligned}
\end{equation}
The wall models for erosion ($\widehat{\Phi}_{sgs}^{src}$) and deposition
($\widehat{\Phi}_{sgs}^{dep}$) are modified for the cut faces, which are not necessarily
perpendicular to the gravity vector; these will be discussed below. Here
$S^\Gamma$ denotes the dimensional area of the cut face. A schematic diagram of
the face areas and cut face in a cut cell can be found in
Fig.~\ref{fig:cut_cell}. For the details of how the geometric quantities (face
area fractions, cell volume fractions, cut face area, and cut face normal
vector) required for the cut cell method are calculated from the level set
function $\varphi$, the reader is referred to Appendix~\ref{sec:calc_cut_cell}.

\subsubsection{Wall Models}

In the Cartesian cut cell framework, the wall models for particle erosion and deposition must be 
modified to account for the possibility of an inclined surface (i.e. where the surface is no longer
perpendicular to the gravity vector). In this case, the deposition model \eqref{eq:pcon_dep} becomes:
\begin{equation}\label{eq:pcon_dep_incline}
  \widehat{\Phi}^{dep}_{sgs} = - (\cos \beta) {w}_s \widetilde{C} \left[ 1 - \left( \frac{h}{z_r} \right)^{-\widehat{\gamma}} \right]^{-1} \text{.}
\end{equation}
where $\beta = \cos^{-1}(\bm{e_3}\cdot\bm{\widehat{n}})$ is slope angle of the cut face,
and $\widehat{\gamma} = {{Sc}_T w_s}/{\kappa \widehat{u}_*}$ is the Rouse
number based on the local shear stress $\widehat{u}_* =
\sqrt{-\widehat{\tau}_{13}/\rho}$, which is calculated 
according to \eqref{eq:local_shear_stress}.  Note that one can recover the
standard wall model for particle deposition over a flat surface
\eqref{eq:pcon_dep} when $\beta = 0$. Conversely, when $\beta = \pi/2$ (a
vertical wall), $\widehat{\Phi}_{sgs}^{dep} = 0$. Here we only consider
deposition due to gravitational settling, and neglect the contribution from
impaction, which is expected to be negligible for our present simulations; this
assumption will be discussed more below.

Similarly, the erosion model on a cut face is given as:
\begin{equation}\label{eq:pcon_src_cut_cell}
  \widehat{\Phi}^{src}_{sgs} = ( \cos \beta) \widehat{w}_s C_r \left( \frac{h}{z_r} \right)^{-\widehat{\gamma}} \left[ 1 - \left( \frac{h}{z_r} \right)^{-\widehat{\gamma}}  \right]^{-1} 
\end{equation}
where we once again use the wall-relative versions of $u_*$ and $\gamma$.  
In the present article, we consider only the case of
deposition, and do not let particles erode once they deposit. A
parameterization of the saltation layer through the wall model will be
addressed in a future study.

\subsubsection{Small Cell Treatment}

One well-known challenge associated with Cartesian cut cell methods is the need
to reduce the timestep in order to satisfy the local Courant-Friedrichs-Lewy
(CFL) condition in cut cells with small volume fractions. For an explicit time
advancement scheme as we adopt here, the timestep must satisfy the condition $C
= {u \Delta t}/{\Delta \ell} < 1$ to avoid numerical instabilities, that is,
the timestep $\Delta t$ must be less than the characteristic convection time
$\Delta \ell  / u$. For cut cells with small volume fractions, the
characteristic cell lengthscale $\Delta \ell$ will be much smaller than the
grid spacing ($\Delta x$, $\Delta y$, or $\Delta z$), thus requiring a much
smaller timestep to ensure the CFL condition is satisfied.

While a number of
approaches have been proposed to circumvent this issue (allowing investigators
to use the same timestep as for the regular Cartesian grid) including cell
linking
\citep[e.g.][]{kirkpatrick2003representation}, cell merging
\citep[e.g.][]{ye1999accurate}, or blended approaches
\citep[e.g.][]{hartmann2008adaptive}, we use the conservative mixing procedure
proposed by \citet{hu2006conservative} and extended to three-dimensional
simulations by \citet{meyer2010conservative}. The conservative mixing procedure
is applied only in small cells (i.e. where $\alpha < 0.5$), which are
identified during the geometric calculations. After calculating the RHS of
\eqref{eq:pcon_cut_cell}, and AB2 time advancement, the particle
concentration in a small cell is mixed with the seven neighboring cells in the
direction of the cut face normal vector $\bm{n^\Gamma}$.  The conservative
mixing procedure conserves mass, and allows us to use the normal timestep we
would employ for a given Cartesian grid. A summary of our implementation of the
conservative mixing model can be found in Appendix \ref{sec:conserv_mix}.

\subsubsection{Velocity Interpolation}

At each timestep, the velocity field calculated from the LES must be
interpolated to the finite volume faces in order to calculate the advective and
diffusive fluxes on the faces of each control volume.  One important constraint on the
interpolation method is the requirement that the interpolated velocity field
remain divergence-free; this is critical in order to conserve mass in the
particle phase. The conservative interpolation method proposed by
\citet{chamecki2008hybrid} does not guarantee that the interpolated velocity
field will be divergence-free in cut cells. In fact, the discrete version of
the continuity equation that must be satisfied in cut cells (where $0 < \alpha
<1$) must include the face area fractions in the discrete divergence operator,
i.e.
\begin{multline}\label{eq:continuity_cut_cell}
  \nabla \cdot \bm{u_{int}} = 
  \frac{1}{\Delta x}[A^x_{i+1/2} U_{i+1/2} - A^x_{i-1/2} U_{i-1/2}] +  \\ 
  \frac{1}{\Delta y}[A^y_{j+1/2} V_{j+1/2} - A^y_{j-1/2} V_{j-1/2}] +  \\
  \frac{1}{\Delta z}[A^z_{k+1/2} W_{k+1/2} - A^z_{k-1/2} W_{k-1/2}] = 0 \text{,}
\end{multline}
where $\bm{u_{int}}$ is the interpolated velocity. For brevity of notation,
unnecessary indicates are omitted from \eqref{eq:continuity_cut_cell} and the
following discussion, i.e.~we will write $A_{i+1/2}$ rather than
$A_{i+1/2,j,k}$, etc.

In order to enforce the divergence-free condition in the cut cells,
we first calculate an intermediate interpolated velocity field using the
conservative interpolation method proposed by \citet{chamecki2008hybrid}. Then,
we remove the finite divergence using a projection method. We first solve the
Poisson equation
\begin{equation}\label{eq:poisson}
  \nabla^2 \psi = \nabla \cdot \bm{u^\star_{int}}
\end{equation}
where $\bm{u^\star_{int}}$ is the intermediate interpolated velocity with
finite divergence in cut cells, then the divergence-free interpolated velocity
field that satisfies \eqref{eq:continuity_cut_cell} is then calculated via
\begin{equation}
  \bm{u_{int}} = \bm{u_{int}^\star} - \nabla \psi \text{.}
\end{equation}
Note that in \eqref{eq:poisson}, the discrete divergence and Laplace operators
must include the face area fractions in the cut cell framework. The right hand
side of \eqref{eq:poisson} is discretized as in \eqref{eq:continuity_cut_cell},
and the Laplacian is discretized using second-order centered differences:
\begin{multline}\label{eq:discrete_laplace}
  \nabla^2 \psi = \\ 
  \frac{1}{(\Delta x)^2} [A^x_{i+1/2} \psi_{i+1} - (A^x_{i+1/2}+A^x_{i-1/2})\psi_{i} + A^x_{i-1/2}\psi_{i-1} ] +  \\
  \frac{1}{(\Delta y)^2} [A^y_{j+1/2} \psi_{j+1} - (A^y_{j+1/2}+A^y_{j-1/2})\psi_{j} + A^y_{j-1/2}\psi_{j-1} ] + \\
  \frac{1}{(\Delta z)^2} [A^z_{k+1/2} \psi_{k+1} - (A^z_{k+1/2}+A^z_{k-1/2})\psi_{k} + A^z_{k-1/2}\psi_{k-1} ]  \text{.}
\end{multline}
Note that one recovers the standard stencils for second-order centered
differences in \eqref{eq:continuity_cut_cell} and \eqref{eq:discrete_laplace}
in the case of cells fully in the fluid, where all of the face area fractions
are identically unity ($A^x = A^y = A^z = 1$). The discrete version of
\eqref{eq:poisson} is solved iteratively at each timestep using the BiCGSTAB(2)
algorithm \citep{sleijpen1993bicgstab} as described in
\citet{van2003iterative}.

\section{Comparison with Wind Tunnel Data}\label{sec:validation}

In order to validate the proposed numerical model, we designed simulations to
compare our LES results with an experimental study of particle deposition over
topography conducted in a wind tunnel \citep{goossens2006aeolian}.  The
detailed deposition measurements in this study make it an ideal test case to
validate our LES model.  A summary of the wind tunnel experiments of
\citet{goossens2006aeolian} can be found in Sec.~\ref{sec:goossens_data}; a
description of the simulations used for the validation cases and a comparison
with the experimental data can be found in Sec.~\ref{sec:les_validation}.

\subsection{Wind tunnel experiments of \citet{goossens2006aeolian}}\label{sec:goossens_data}

The particle deposition experiments of \citet{goossens2006aeolian} were
conducted in a closed-return wind tunnel with a test section of dimensions 7.6
m long, 1.2 m wide, and 0.60 m high. Dust was released in the return section of
the wind tunnel, and three separate concave-convex symmetric hills were
constructed out of zinc plates in order to study the deposition patterns over
these topographic features.  For our validation case described below in
Sec.~\ref{sec:les_validation}, we compare our LES results to deposition onto
the hill (``hill 2'') having dimensions of height $H = 0.03$ m and length $L =
0.06$ m.  The turbulent boundary layer (as measured 0.5 m upstream of each
hill) was characterized by free stream velocity of $U_{\infty} = 1.72$ m
s$^{-1}$, friction velocity $u_* = 0.06$ m s$^{-1}$, and roughness length $z_0
= 1\times 10^{-5}$ m (i.e. the surface was hydrodynamically smooth). The
Reynolds number based on hill height (sometimes called the ``roughness Reynolds
number'') for hill 2 was ${Re}_h = {u_* h}/\nu = 123$;
\citet{cermak1984physical} found that Reynolds-number independence for wind
tunnel studies of flows over topography was achieved for ${Re}_h > 70$;
therefore the Reynolds number criterion was found to be satisfied. Each hill
was placed in the wind tunnel with a fetch of 5.05 m from the beginning of the
test section to the beginning of the hill.

Dust particles were released in the return section of the wind tunnel at a rate
of 13 kg h$^{-1}$; the dust used in the experiment was prepared from calcareous
loam with a mean particle diameter of $\overline{d}_p = 42\, \mu$m, maximum
particle size of $d_p^{max} = 104\, \mu$m, and a mass density of $\rho_p =
2650$ kg m$^{-3}$.  \citet{goossens2006aeolian} notes that the resuspension of
particles was expected to be small due to the small friction velocity.
Deposition was measured at 0.01 m increments at 250 locations in the
longitudinal direction and as a function of particle size for nine grain size
classes.  Each particle release experiment was 12 minutes in duration.

\subsection{Large eddy simulations}\label{sec:les_validation}

\begin{table*}
  \begin{center}
  \begin{tabular}{c c c c} \hline
  Quantity & Symbol & Coarse Resolution & Fine Resolution \\ \hline
  Domain [m] & $L_x \times L_y \times L_z$ & $1.28 \times 0.64 \times 0.24$ & $1.28 \times 0.64 \times 0.24$ \\
  Number of gridpoints [-] & $N_x \times N_y \times N_z$ & $64\times32\times48$ & $128\times64\times96$ \\
  Grid spacing [m] & $\Delta x \times \Delta y \times \Delta z$ & $0.02^2 \times0.005$ & $0.01^2 \times0.0025$ \\
  Timestep, Smag [s] & $\Delta t$ & $1.5\times10^{-3}$ & $7.5\times10^{-4}$ \\
  Timestep, Dynamic [s] & $\Delta t$ & $1.5\times10^{-3}$ & $4.0\times10^{-4}$\\
  Timestep, LASD [s] & $\Delta t$ & $1.5\times10^{-3}$ & $4.0\times10^{-4}$ \\
  Friction velocity [m s$^{-1}$] & $u_*$ & 0.06 & 0.05 \\
  Characteristic timescale [s] & $T_\ell = L_z/u_*$ & 4.0 & 4.8 \\ 
  Duration of particle release [-] & $n T_\ell$ & $n=30$ & $n=25$  \\
  Roughness length [m] & $z_0$ & $1.0\times10^{-5}$ &  $1.0\times10^{-5}$ \\
  Settling velocity [m s$^{-1}$] & $w_s$ & 0.10 & 0.10 \\
  SGS Schmidt number [-] & $\text{Sc}_{sgs}$ & 1.0 & 1.0 \\
  Hill height [m] & $H$ & 0.03 & 0.03 \\
  Hill length [m] & $L$ & 0.06 & 0.06 \\ 
  Hill centerline location [m] & $x_0$ & 0.64 & 0.64 \\ 
  Stokes number [-] & ${St}_H = {\tau_p u_*}/H$ & 0.02 & 0.017 \\ \hline 
  \end{tabular}
  \end{center}
  \caption{Simulation properties for validation cases.}
  \label{tab:validation_les}
\end{table*}

Large eddy simulations with heavy particle deposition over surface topography
were conducted in order to validate the proposed numerical model with the wind
tunnel data of \citet{goossens2006aeolian}. Neutral flow simulations were
driven by a constant pressure gradient force, and a two-dimensional sinusoidal
hill (spanning the entire domain in the $y$-direction),
described by 
\begin{equation}\label{eq:sin_hill} h(x) = H \cos^2 \left(
    \frac{\pi (x-x_0)}{2 L} \right) 
\end{equation} 
was included in the simulations, where $H$ is the hill height, $L$ is the hill
length, and $x_0$ is the streamwise coordinate of the hill crest.  Simulations
were run at a coarse ($64 \times 32 \times 48$) and fine ($128 \times 64 \times
96$) resolution, and using three subgrid models---static Smagorinsky
\citep{smagorinsky1963general} with $c_s = 0.10$ and wall damping included
\citep[e.g.][]{mason1994large}, the plane-averaged dynamic model
\citep{lilly1992proposed}, and the Lagrangian-averaged scale-dependent
dynamic Smagorinsky model \citep{bouzeid2005scale} in order to compare the
effects of grid resolution and SGS model on the observed deposition patterns.

Particles were released from an area source in the $y$-$z$ plane near the
leading edge of the domain (at node $jx=3$), with a release rate of $Q = 3.61$
g s$^{-1}$ (13 kg h$^{-1}$). Simulation results were compared to the data for
the 31--41 $\mu$m size class from \citet{goossens2006aeolian}. (Note that
\citet{goossens2006aeolian} found the normalized particle deposition patterns
did not vary greatly for the different particle size classes). Periodic
boundary conditions were employed for the concentration field in the
$y$-direction and inflow/outflow conditions in the $x$-direction.

Particle settling velocity was calculated from Stokes' drag law
\begin{equation}\label{eq:stokes_drag}
w_s = \frac{\rho_p g d_p^2}{18 \mu_{air}}, 
\end{equation}
which is valid for small particle Reynolds numbers ($\text{Re}_p = {w_s
  d_p}/{\nu_{air}}$), and where $\rho_p$ is particle density and $\mu_{air}$ is
the dynamic viscosity of air \citep[e.g.][]{pruppacher1998microphysics,
  lamb2011physics}.  For 36 $\mu$m particles, $\text{Re}_p \simeq 0.24$ and
$w_s = 0.10$ m s$^{-1}$. Note that using a semi-empirical formula \citep[][pp.
388-390]{lamb2011physics} to estimate $w_s$ (the typical approach for large
$\text{Re}_p$) yielded a similar estimate of settling velocity.

Simulations were spun up on a coarse grid with no particles for approximately
30 large-eddy turnover times, where $T_\ell = L_z/u_*$. Particles then were
released for 25-30 $T_\ell$ (2 minutes physical time). In the case of the
simulations on the fine grid, this was done after trilinear interpolation of
the velocity field to the fine grid and evolving the velocity field for several
$T_\ell$. A summary of the properties of the simulations for the validation
cases can be found in Table~\ref{tab:validation_les}.

One well-known property of heavy particles in turbulent flow is the fact that
particles will not follow fluid streamlines exactly due to their inertia.  This
effect can be quantified by the Stokes number, a dimensionless number defined
as ${St} = \tau_p/\tau_f$, where $\tau_p$ is once again the characteristic
particle timescale, and $\tau_f$ is a characteristic fluid timescale. In the
${St}\rightarrow\infty$ limit, the particles will settle without feeling the
effect of the fluid; when ${St}\rightarrow0$, the particles are passive tracers
that follow the motion of the fluid exactly. The trajectory-crossing effect
becomes significant for ${St}\sim\mathcal{O}(1)$.

Two variations of the Stokes number are relevant for the results presented
here. The first is the grid Stokes number ${St}_\Delta = \tau_p/\tau_\Delta$,
where $\tau_\Delta$ is the timescale of the smallest turbulent eddies resolved
in an LES. Note that $\tau_\Delta$ can be estimated via \citep[e.g.][]{yang2016large}
\begin{equation}\label{eq:tau_delta}
  \tau_\Delta \sim T_\ell \left(\frac{\Delta}{\ell}\right)^{2/3} \sim \frac{L_z}{u_*}\left(\frac{\Delta}{L_z}\right)^{2/3}
\end{equation}
where $\ell$ and $T_\ell$ are respectively the integral length and time scales.
As discussed by \cite{shotorban2007eulerian}, the equilibrium Eulerian approach
(where the particle velocity can be explicitly computed from the fluid velocity
field as we do in the present work) has been found to be valid up to
${St}_\Delta \lesssim 0.5$ from comparison with direct numerical simulation (DNS)
results. For the present study, ${St}_\Delta \sim \mathcal{O}(0.01)$ for the
validation cases presented in Sec.~\ref{sec:validation} and ${St}_\Delta \sim
\mathcal{O}(1\times10^{-4})$ for the snow deposition case study presented in
Sec.~\ref{sec:snow_deposition}, demonstrating that it is reasonable to adopt
the equilibrium Eulerian approach here.

The second relevant parameter is the Stokes number based on hill height and
friction velocity, i.e.  ${St}_H = \tau_p u_*/H$, which is the relevant
parameter for characterizing the extent to which the trajectory-crossing effect
enhances particle deposition.  Previous studies of the deposition of inertial
particles onto obstacles \citep[e.g.][]{may1967impaction, aylor1982modeling,
  moran2013understanding} have used a Stokes number based on the characteristic
velocity scale of the flow and obstacle lengthscale (e.g. ${St}={\tau_p
  u_0}/{\ell_0}$) to characterize the role of particle inertia. For the present
study, we define this Stokes number in terms of hill height and friction
velocity; values for the simulations considered here are reported in
Tables~\ref{tab:validation_les} and \ref{tab:snow_les}. 
Based on the fit by \cite{aylor1982modeling} to the data of \cite{may1967impaction} for
the impaction of inertial particles onto cylinders in crossflow, i.e.
\begin{equation}
E_I = 0.86 \, (1 + 0.442 \,{St}^{-1.967})^{-1},
\end{equation}
the impaction efficiency for the largest value of ${St}_H$ considered here
(${St}_H = 0.02$) is negligible, i.e. $E_I \sim \mathcal{O}(1\times10^{-3})$.
(Note that impaction efficiency, where $0 \leq E_I \leq 1$, is simply the
fraction of particles that will impact onto an obstacle due to the
trajectory-crossing effect).  Thus neglecting the contribution from impaction
in the wall model for particle deposition \eqref{eq:pcon_dep_incline} is a
reasonable assumption for the range of ${St}_H$ considered here.

\begin{figure*}[htbp]
  \begin{center}
    \includegraphics[scale=1.0]{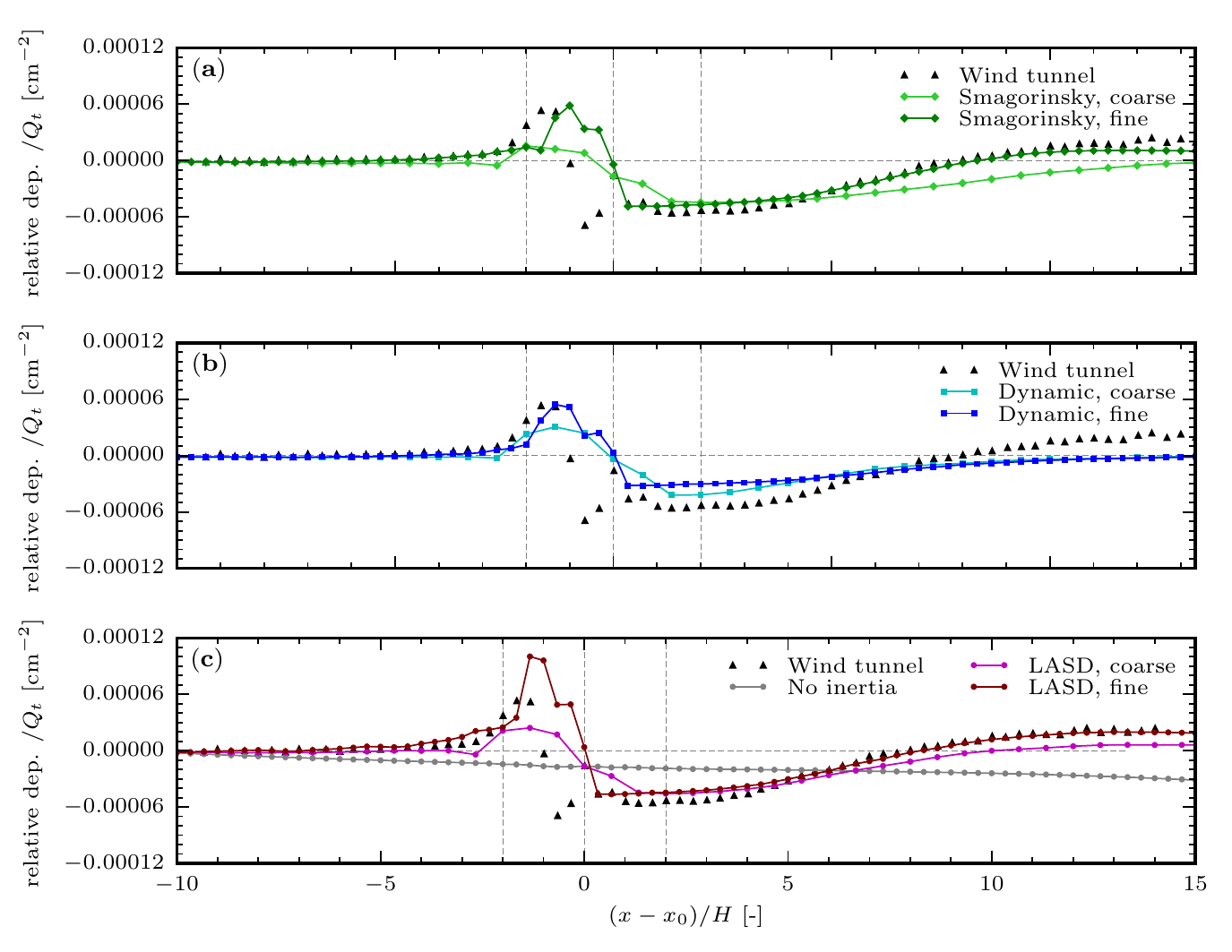}
  \end{center}
  \caption{Particle deposition over topography from the LES at coarse
    ($64\times 32 \times 48$) and fine ($128 \times 64\times 96$) resolution,
    together with the wind tunnel data of \citet{goossens2006aeolian}. Here
    relative deposition, normalized by the total mass of particles released
    ($Q_t$) is plotted as a function of distance relative to the streamwise
    location of the hill crest ($x_0$), normalized by hill height ($H$). (a)
    Static Smagorinsky model with $c_s = 0.1$ (b) plane-average dynamic model
    (c) Lagrangian scale-dependent dynamic model.}
  \label{fig:validation}
\end{figure*}

\subsection{Particle deposition}\label{sec:particle_deposition}

A comparison between the LES and the wind tunnel data can be found in
Fig.~\ref{fig:validation}. Results from the static Smagorinsky SGS model can be
found in panel (a), the plane-averaged dynamic model in panel (b), and the
Lagrangian scale-dependent dynamic model in panel (c).  In each panel, the
relative deposition is normalized by the total mass of particles released
($Q_t$) and plotted as a function of normalized distance from the streamwise
location of the hill crest ($x_0$).  Following \citet{goossens2006aeolian}, we
define the relative deposition as the deposition with the mean upwind of the
hill (from $(x-x_0) \in [-50,10]$ cm) removed. One can see from
Fig.~\ref{fig:validation} that the relative deposition from the wind tunnel
data increases on the windward side of the hill, then decreases rapidly on the
leeward side. Downwind of the hill, the relative deposition is initially
negative, but increases downstream of the hill, and becomes positive at
$(x-x_0) \approx 7 H$ (i.e. deposition is larger here than what is found upwind
of the hill).  This same qualitative pattern of particle deposition over hills
has been observed both in the field and in experimental studies at various
scales \citep{goossens2006aeolian}.

On the coarse ($64\times32\times48$) grid, all three SGS models recover the
correct trends in the deposition pattern, where the maximum of deposition is
found on the windward side of the hill, the minimum on the leeward side, and
deposition increases going downwind of the hill in the wake. However, all three
SGS models underpredict the peak of deposition on the windward side of the
hill, and overpredict deposition on the leeward side.

On the fine ($128\times64\times96$) grid, the Smagorinsky model, displayed in
Fig.~\ref{fig:validation}(a) has a reasonable prediction of the deposition
patterns compared to the wind tunnel data. On the windward side of the hill,
the peak of the deposition is slightly displaced, occurring closer to the hill
crest than in the experimental data. However, the Smagorinsky model is in good
agreement with the wind tunnel data on the leeward side of the hill, and in the
wake, only beginning to deviate slightly around $x-x_0 \approx 10 H$.

The plane-averaged dynamic model, as seen in Fig.~\ref{fig:validation}(b), has
a more accurate prediction of the peak of deposition on the windward side of
the hill than Smagorinsky, although it also slightly displaces the peak of
deposition toward the crest of the hill. However, here the dynamic model does
not capture the behavior of the deposition in the wake as well as Smagorinsky,
overpredicting deposition in the near wake, and underpredicting deposition in
the far wake. 

The deposition pattern from the LASD model, visible in
Fig.~\ref{fig:validation}(c), begins overpredicting deposition before the start
of the hill (e.g.~around $x-x_0 \approx 4 H$). The LASD model significantly
overpredicts the peak of deposition on the windward side of the hill. However,
the LASD model successfully captures the deposition in the wake, here
performing better than both the dynamic and Smagorinsky models. An additional
simulation using the LASD model was run on the fine grid, but without the model
for particle inertia (i.e. by omitting the $-\tau_p \widetilde{\bm{a}}$ term in
the particle advection velocity \eqref{eq:advection_vel}). One can see that the
curve for the no inertia case, displayed in Fig.~\ref{fig:validation}(c) fails
to recover the correct deposition patterns, demonstrating that accounting for
particle inertia is critical when simulating the deposition of heavy particles
over complex topography. The effects of particle inertia on the concentration
field and deposition patterns will be discussed further in
Sec.~\ref{sec:snow_deposition}.

The overprediction of deposition on the windward side of the hill that is
visible in Fig.~\ref{fig:validation}(c) for the LASD model on the fine grid is
likely related to the fact that the LASD model uses information inside the IBM
interface (e.g.~the velocity field that has been set to zero, then polynomial
smoothing applied) when determining a value of the Smagorinsky coefficient
$c_s$ for nodes near the boundary; note that the LASD model uses test filters
at scale $2\Delta$ and $4\Delta$ to determine $c_s$; thus values of $c_s$ (and
consequently the SGS scalar diffusivity $K_c^{sgs} = (c_s \Delta)^2
|\widetilde{\bm{S}}|/\text{Sc}_{sgs}$) on nodes near the IBM interface will be
influenced by the nonphysical values of velocity within the immersed boundary
surface.  Nevertheless, on the fine grid the three SGS models considered are in
reasonable agreement with the experimental data.

\section{Case Study of Snow Deposition}\label{sec:snow_deposition}

In this section, we use the proposed numerical model to investigate the
deposition of snow over idealized surface topography. As discussed earlier,
most previous studies of snow transport have followed the RANS approach, where
all scales of turbulence are parameterized \citep{gauer2001numerical,
  schneiderbauer2011atmospheric}, or have employed Lagrangian particle models
and focused on drifting at small spatial scales \citep{zwaaftink2014modelling}.
In a recent article, \cite{lehning2008inhomogeneous} focused on the extent to
which topography modifies mean wind fields in complex terrain, thereby
modifying the deposition velocity, leading to preferential deposition of snow.
However, the effect of the spatially heterogeneous flow fields induced by
surface topography and particle inertia on deposition patterns in the context
of snow transport has not, to our knowledge, been explored using LES in an
Eulerian framework.

\subsection{Large eddy simulations}\label{sec:snow_les}

In order to investigate the effects of topography on snow deposition patterns
(an important question for hydrological modeling, alpine ecology, and avalanche
forecasting), we simulate the deposition of fresh snow, where particles are
released from an area source in the $x$-$y$ plane near the top of the domain.
Idealized surface topography---here a two-dimensional sinusoidal hill extending
across the entire domain in the $y$-direction, described by
\eqref{eq:sin_hill}---was included in the simulation.  Here the roughness
Reynolds number is approximately ${Re}_h \approx 2.6 \times 10^5$; thus
${Re}_h$ independence is found to be satisfied.  While snow deposition and
transport in the real world can included a number of complexities, including
saltation, non-neutral thermal stratification, erosion of snow once it is
deposited, an evolving surface due to erosion and/or deposition, sublimation of
blowing snow, fracturing and sintering of ice particles, and polydisperse
particles, we here perform idealized simulations.  We restrict ourselves to
monodisperse particles (with a constant settling velocity $w_s$), neutral
stratification, and a fixed surface height. Once particles deposit, they are
not allowed to be resuspended.

Parameters for this simulation can be found in Table~\ref{tab:snow_les}. The
velocity field for these simulations was spun up on a coarse ($64\times 32
\times 48$) grid, then interpolated to the finer ($128 \times 64 \times 96$)
grid and evolved for several $T_\ell$ before beginning the particle release.
The plane-averaged dynamic SGS model \citep{lilly1992proposed} was employed.
Both the spinup and particle release were one hour physical time in duration
(approximately 7.6 $T_\ell$).  We used a settling velocity of $w_s = 0.1$ m
s$^{-1}$, which is within the range of values that have been reported for fresh
snow particles \citep[e.g.][]{kajikawa1972measurement,
  lehning2008inhomogeneous}. While particle deposition patterns are expected to
be a function of the Stokes number $\text{St}_H$, an exploration of the full
parameter space is beyond the scope of the present article. 

\begin{table*}
  \begin{center}
  \begin{tabular}{c c c} \hline
  Quantity & Symbol & Value \\ \hline
  Domain [m] & $L_x \times L_y \times L_z$ & $640 \times 320 \times 96$  \\
  Number of gridpoints [-] & $N_x \times N_y \times N_z$ & $128\times64\times96$ \\
  Grid spacing [m] & $\Delta x \times \Delta y \times \Delta z$ & $5\times 5 \times1$ \\
  Timestep [s] & $\Delta t$ & 0.025  \\
  Friction velocity [m s$^{-1}$] & $u_*$ & 0.20 \\
  Characteristic timescale [s] & $T_\ell = L_z/u_*$ & 473 \\
  Duration of particle release [-] & $n T_\ell$ & $n = 7.6$ \\
  Roughness length [m] & $z_0$ & 0.01 \\
  Settling velocity [m s$^{-1}$] & $w_s$ & 0.10 \\
  SGS Schmidt number [-] & $\text{Sc}_{sgs}$ & 1.0 \\
  Release rate [kg s$^{-1}$] & $Q_{src}$ & 145. \\
  Source height [m] & $z_{src}$ & 95 \\
  Hill height [m] & $H$ & 20 \\
  Hill length [m] & $L$ & 40 \\ 
  Hill centerline [m] & $x_0$ & 80 \\ 
  Stokes number [-] & ${St}_H = {\tau_p u_*}/H$ & $1.0\times10^{-4}$ \\ \hline 
  \end{tabular}
  \end{center}
  \caption{Simulation properties for snow deposition case study.}
  \label{tab:snow_les}
\end{table*}

\subsection{Results}\label{sec:snow_results}

In this section we present results of large eddy simulations of snow deposition
onto surface topography. In Sec.~\ref{sec:mass_balance}, we will consider the
mass conservation of the proposed model; instantaneous and time-average
snapshots of the flow will be presented in Sec.~\ref{sec:visualization}.
Deposition patterns will be discussed in Sec.~\ref{sec:deposition}, and the
connection between flow acceleration, particle inertia, and the observed
deposition patterns will be explored in Sec.~\ref{sec:accel}.

\begin{figure*}[htbp]
  \begin{center}
   \includegraphics[scale=1.0]{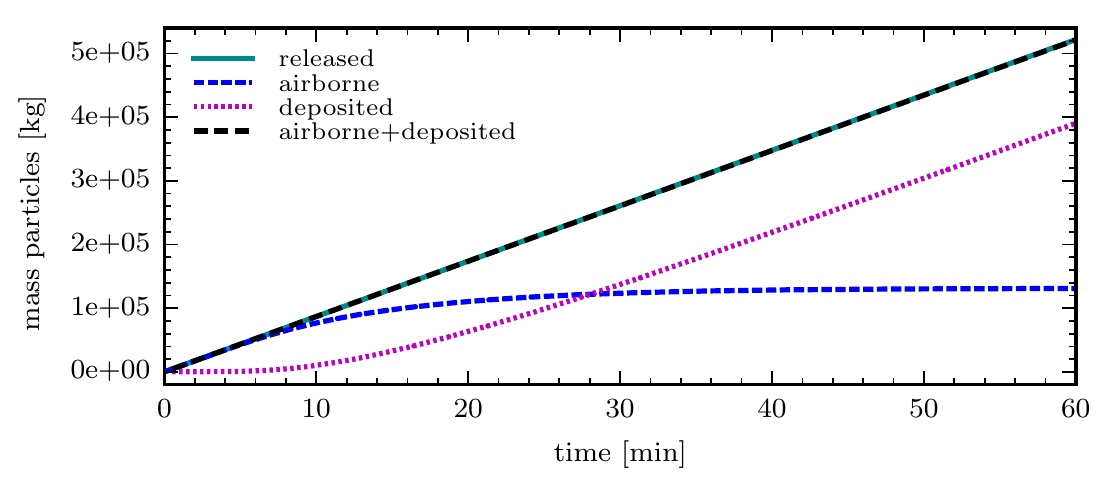}
  \end{center}
  \caption{Mass balance of 1 hour particle release over surface topography, including total mass of 
  particles released, airborne, deposited, and the sum of airborne $+$ deposited.}
  \label{fig:mass_balance}
\end{figure*}

\subsubsection{Mass Conservation}\label{sec:mass_balance}
As discussed earlier in Sec.~\ref{sec:cut_cell}, the motivation for developing
a Cartesian cut cell method is to explicitly enforce mass conservation in the
particle phase. In Fig.~\ref{fig:mass_balance}, we present the global mass
balance for a 1 hour particle release over surface topography. Separate curves
indicate the mass of particles released, airborne, and deposited, and the sum
of airborne $+$ deposited, which should match the release rate if mass
conservation is accurate. Here one can see that the global mass balance reaches
equilibrium after a time of approximately $L_z/w_s \sim 960$ s (or 16 minutes),
which is approximately the time it takes for the first particles released to
begin depositing on the surface. After this, the mass of particles airborne and
the deposition rate are both constant. Note that the mass of particles airborne
$+$ deposited is very close to the total mass released. After a 1 hour particle
release, the residual of the mass balance (i.e. $(\text{released} -
\text{airborne} - \text{deposited})/\text{released}$) is within 0.10\%, which
demonstrates that the proposed numerical model does indeed conserve mass
accurately. 

It must be emphasized that accurate mass conservation is essential for
physically realistic simulations of the transport of heavy particles or passive
scalars over complex terrain. In order to ensure mass conservation, the
treatment at the interface is of central importance; spurious fluxes induced by
truncation errors or non-conservative treatment of quantities at the
fluid-solid interface can deteriorate the quality of both local and global
statistics. As discussed by \cite{mittal2005immersed}, local and global mass
conservation in the presence of an immersed boundary can only be guaranteed in
a finite-volume framework, as we employ in the Cartesian cut cell method
developed here.

\subsubsection{Velocity and concentration fields}\label{sec:visualization}

\begin{figure*}[htbp]
  \begin{center}
    \includegraphics[scale=1.0]{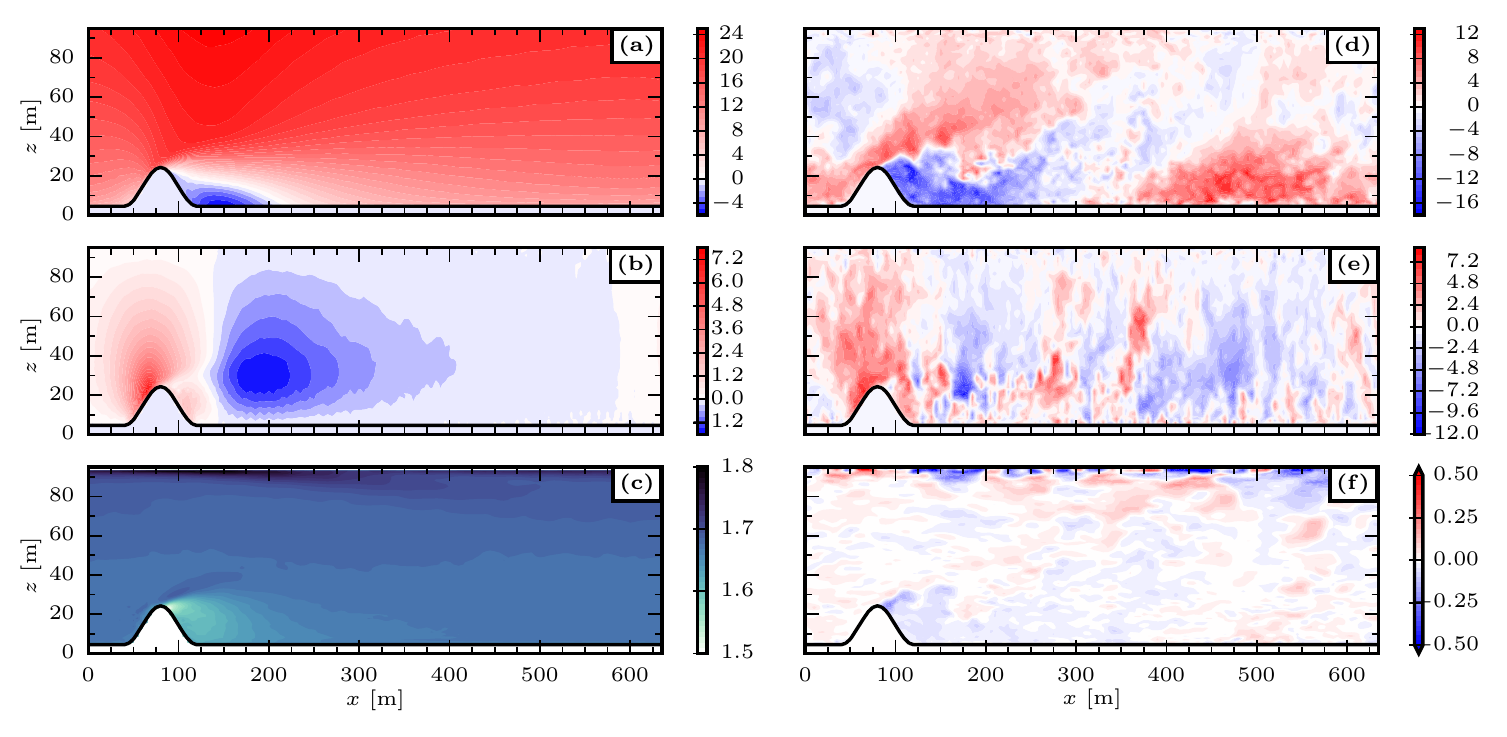}
  \end{center}
  \caption{Visualization of velocity and particle concentration from LES. Time averaged 
    quantities are displayed in panels (a)--(c) and instantaneous fluctuating quantities (with the horizontal mean removed)
    at $y = Ly/2$ are displayed in panels (d)--(f).
    (a) Average streamwise velocity $ U/u_*$ 
    (b) average vertical velocity $ W/u_*$
    (c) average particle concentration $C/c_*$
    (d) fluctuating streamwise velocity ${u}'/u_*$
    (e) fluctuating vertical velocity ${w}'/u_*$
    (f) fluctuating particle concentration ${c}'/c_*$.}
  \label{fig:uwc_inst_avg}
\end{figure*}


A visualization of the instantaneous and averaged velocity and concentration
fields can be found in Fig.~\ref{fig:uwc_inst_avg}.  Here we use the Reynolds
averaging convention, where a total resolved variable from the LES is
decomposed into its mean and fluctuating parts, e.g. $\widetilde{u} = U + u'$.
Fluctuating quantities are calculated by removing the mean value calculated by
averaging in the fluid only. Although this flow cannot be considered
horizontally homogeneous due to the presence of topography, we do this because
our focus is on how topography modifies quantities from their values found in
the corresponding case over statistically homogeneous, flat terrain.  For
simplicity of notation, we shall also omit the tilde ($\widetilde{\,\cdot\,}$)
from filtered variables, unless necessary.  Variables are normalized using the
friction velocity $u_* = \sqrt{-\tau_w/\rho} = (\overline{u'w'}^2)^{1/4}$ and
the concentration scale $c_* = |\overline{w'c'}_0|/u_*$, where
$|\overline{w'c'}_0|$ is the magnitude of the surface concentration flux. The
velocity and concentration scales $u_*$ and $c_*$ are calculated by averaging
just outside immersed boundary interface (i.e.~in the region $0 \leq
\varphi \leq \Delta z$); mean values of concentration, velocity, turbulent
fluxes, etc.~displayed in Figs.~\ref{fig:uwc_inst_avg}--\ref{fig:inertia_terms}
are calculated by averaging in time and in the $y$-direction.

One can see from Fig.~\ref{fig:uwc_inst_avg}(a)-(b) that for this particular
hill geometry, the flow detaches in the wake of the hill and a separation
bubble forms such that $U/u_* < 0$. (Note that flow separation over smooth
two-dimensional ridges is expected to occur for hill slopes of $\theta \gtrsim
18^{\circ}$, where $\theta$ decreases with increasing surface roughness
\citep{kaimal1994atmospheric}). The reattachment point is slightly more than
$2L$ from the end of the hill on the leeward side. Due to the topography, there
is also a region of positive average vertical velocity ($W>0$) on the windward
side of the hill, and average negative vertical velocity ($W<0$) on the leeward
side. A region of low particle concentration is found in the wake of the hill,
and extends slightly beyond the reattachment point. In addition, a ``streamer''
of high concentration extends beyond the hill crest.  Similar patterns are
visible in the instantaneous velocity and concentration fields, displayed in
Fig.~\ref{fig:uwc_inst_avg}(d)--(f).


\subsubsection{Deposition Patterns}\label{sec:deposition}
\begin{figure*}[htb]
  \begin{center}
    \includegraphics[scale=1.0]{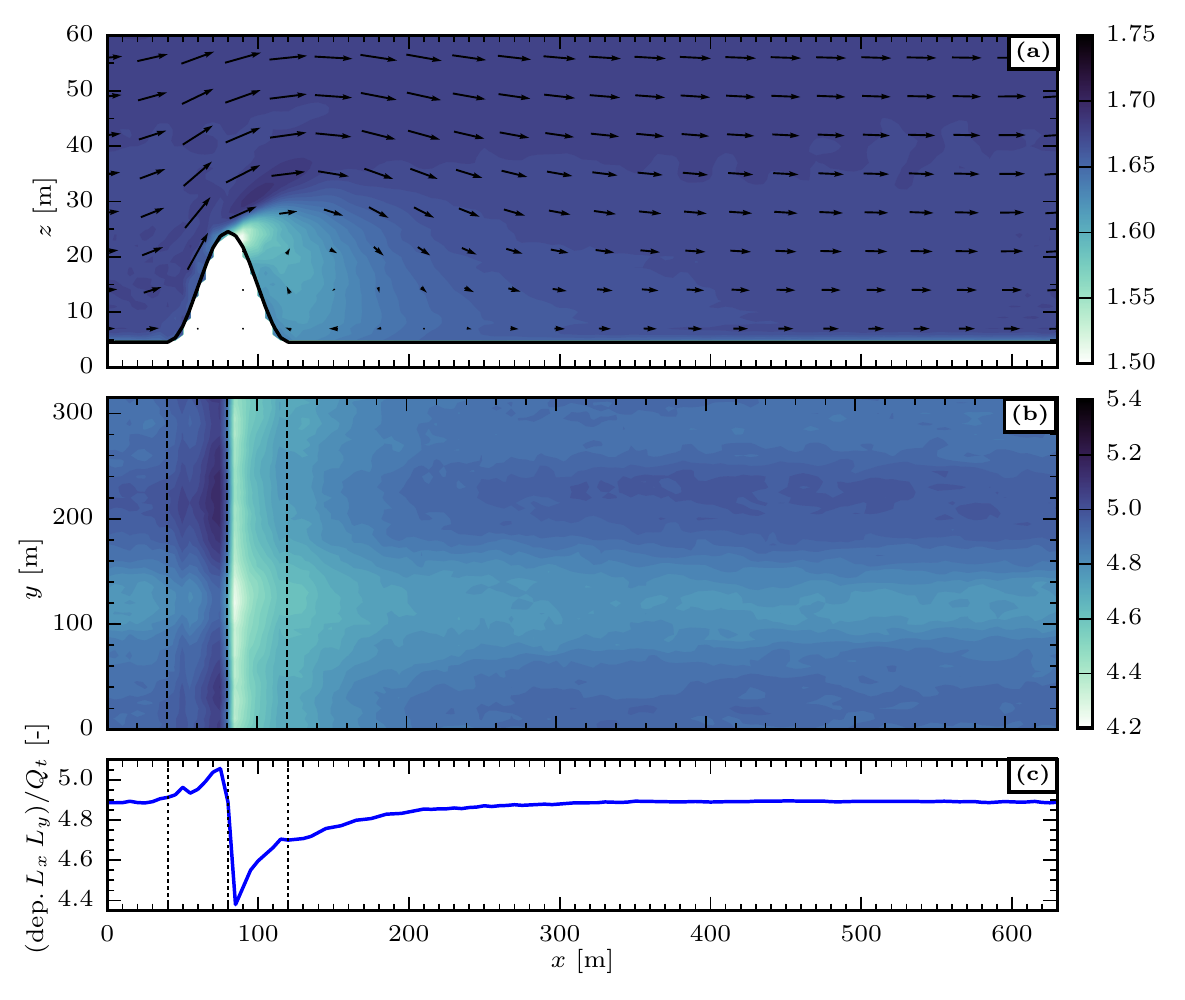}
  \end{center}
  \caption{Average velocity and particle concentration and deposition from area
    source release over two-dimensional hill.  (a) Contours of average particle
    concentration $C/c_*$  in $x$-$z$ plane
    with velocity vectors overlaid (b) nondimensional deposition, i.e.
    $(\text{deposition} \times L_x \times L_y)/Q_t$, in $x$-$y$ plane after 1
    hour particle release (c) nondimensional $y$-averaged deposition as a function of $x$.
  }
  \label{fig:deposition}
\end{figure*}

A plot of the average concentration field in the $x$-$z$ plane with velocity
vectors overlaid can be found in Fig.~\ref{fig:deposition}(a); in panel (b) we
show a plot of deposition patterns from the LES after a 1-hour particle release.
Panel (c) contains a figure of the $y$-averaged deposition after the conclusion of
the particle release. Note that the dashed lines in panels (b)--(c) correspond
to the edges and crest of the sinusoidal hill. We find that the deposition from
the LES increases on the windward side of the hill, decreases sharply on the
leeward side, and reaches a quasi-constant value in the far wake, as seen in
panel (c). The maximum value of deposition (approximately 4-5\% larger than the
upwind value) is found on the windward side of the hill, slightly before the
crest; the minimum occurs on the leeward side just beyond the crest and is
$\sim$9\% lower than the upwind value. This deposition pattern is qualitatively
similar to previous observational and laboratory studies of particle deposition
onto topography \citep[e.g.][]{goossens2006aeolian}. The deposition patterns
can be seen more clearly in the planview, plotted in
Fig.~\ref{fig:deposition}(b). Note that, although the topography and the flow
can be considered to be homogeneous in the $y$-direction, some lateral
variability in the mass of particles deposited does occur downwind of the hill
(e.g. around $x = 200$ to $600$ m).  We found from simulations of particle
deposition onto a flat surface (not shown) with the same grid and domain that
particle deposition varies spatially by no more than $\pm$ 1\% when no
topography is present.  When topography is present, deposition varies in the
spanwise direction by $\sim$8\% in the wake of the hill; this is due to
long-lived high- and low- speed streaks in the streamwise velocity field.

Average contours of the particle concentration field with velocity vectors
overlaid can be found in Fig.~\ref{fig:deposition}(a). Here one can see that
the region of high particle concentration immediately upwind of the hill is
responsible for the enhanced deposition on the windward side of the hill;
likewise, the low deposition in the wake is due to the lower particle
concentrations. From Fig.~\ref{fig:deposition} one can see that the
$y$-averaged deposition does not vary significantly in the streamwise direction
after about $x = 300$ m; this is where $C$ becomes quasi-constant in the
streamwise direction and occurs approximately $80$ m after the reattachment
point. Note, however, that the reattachment point and the location where $C$
recovers to a quasi-constant value do not coincide due to particle inertia,
which leads to relative velocities between the particles and fluid, which will
be discussed more below.

\subsubsection{Particle Inertia and Acceleration}\label{sec:accel}

The region of increased particle concentration beginning on the windward side
of the hill and extending beyond the crest and the subsequent increase in
deposition can be understood by noting that heavy particles will not follow the
fluid flow field exactly due to their inertia. This ``preferential
concentration'' effect is well-known in the multiphase flow literature
\citep[e.g.][]{squires1991preferential, wang1993settling,
  eaton1994preferential}; the extent to which it occurs is a function of the
Stokes number.  However, many of these studies focused on preferential
concentration in homogeneous isotropic turbulence. We here consider the effects
of particle inertia on the concentration field in wall-bounded turbulent shear
flows over complex topography and the implications for particle deposition.

\begin{figure*}[htbp]
  \begin{center}
    \includegraphics[scale=1]{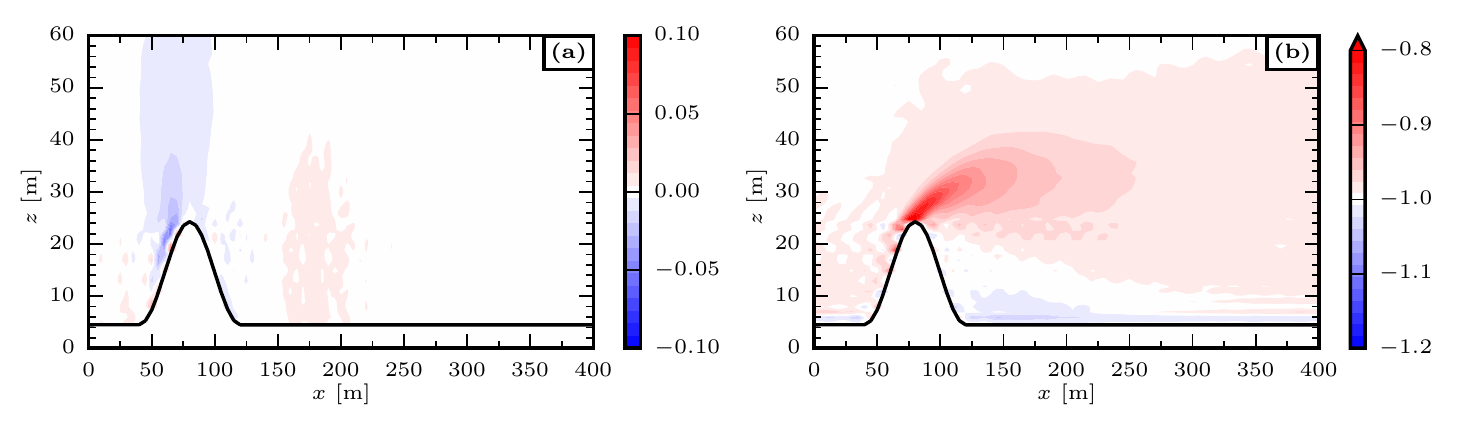}
  \end{center}
  \caption{Relative velocity between particles and fluid $\bm{v^{rel}} =
    \widetilde{\bm{v}}_p - \widetilde{\bm{u}} = \tau_p \bm{g} - \tau_p
    \widetilde{\bm{a}}$, normalized by the background settling velocity $w_s$.
    (a) $x$-component of relative velocity, $v_x^{rel}/w_s = -\tau_p a_x/w_s$
    (b) $z$-component of relative velocity, $v_z^{rel}/w_s = -(\tau_p a_z +
    w_s)/w_s$.}
  \label{fig:rel_vel}
\end{figure*}

From the expression for the
particle advection velocity \eqref{eq:advection_vel}, one can write the relative velocity 
between the particles and fluid as:
\begin{equation}\label{eq:rel_veloc}
  \bm{\widetilde{v}}_{{rel}} = \widetilde{\bm{v}}_p - \widetilde{\bm{u}} = 
  \tau_p \bm{g} - \tau_p \widetilde{\bm{a}}\text{,}
\end{equation}
demonstrating that relative velocities between the particles and fluid occur
due to a combination of gravitational settling and particle inertia.

The time- and laterally-averaged $x$- and $z$-components of relative velocity
are plotted in Fig.~\ref{fig:rel_vel} in panels (a) and (b), respectively.  One
can see that the $x$-component of relative velocity is negative on the windward
side of the hill due to the positive streamwise acceleration $a_x > 0$ in this
region. Here particles cross fluid streamlines, leading to the ``streamer'' of
high concentration observed in Fig.~\ref{fig:deposition}(a) and the
subsequent increase in deposition on the windward side of the hill. In this
region, $v_x^{rel}$ is negative, and has a magnitude of $\sim$5-6\% of $w_s$.
A region of positive $v_x^{rel}$ is found in the wake of the hill where the
flow decelerates. Here the magnitude of $v_x^{rel}$ is $\sim$4-5\% of $w_s$.
In Fig.~\ref{fig:rel_vel}(b), one can see that far away from the hill, the
$z$-component of relative velocity is due strictly to the constant
gravitational settling velocity. However, downstream and above the crest of the
hill, the vertical component of relative velocity is reduced; here $v_z^{rel}$
is as small as $\sim$80\% of $w_s$. This region corresponds to where low
concentration is found in the wake, as once can see in
Fig.~\ref{fig:uwc_inst_avg}, where deposition is at its minimum.  

When \cite{lehning2008inhomogeneous} introduced the concept of preferential
deposition of snow in mountains, they suggested that more snow may be deposited
in the lee of a steep alpine ridge. The study of \cite{mott2010meteorological}
obtained similar results for the same ridge, while \cite{mott2010understanding}
found more preferential deposition on the windward side of steep slopes for a
different experimental area (their Fig.~14a). Our results clearly suggest that
lee slopes may receive a minimum of precipitation.  This is in partial
contradiction to the earlier ``simple ridge'' cases.  The combination of
Advanced Regional Prediction System (ARPS) RANS wind simulations with Alpine3D
deposition modeling as applied in all three previous investigations does not
properly account for particle inertia, which may explain some of the
differences. On the other hand, \cite{mott2014orographic} have recently shown
that higher concentrations are often found downwind of ridges, at least higher
up in the atmosphere.  Further studies are therefore required to determine the
dependence of preferential deposition patterns on the wind fields and
topography (which can be expressed in terms of the Stokes number $\text{St}_H$)
as well as other factors (e.g. atmospheric stability).  The results for the
relative velocities and deposition patterns presented here therefore should not
be taken as general; a more comprehensive investigation of the dependence of
preferential deposition on ${St}_H$ and other parameters will be addressed in
future work.

In order to investigate the role of mean vs.~fluctuating inertia for the
preferential concentration of particles, one can Reynolds average the inertia
terms that appear in the particle conservation equation
\eqref{eq:pcon}, i.e.  
\begin{equation}\label{eq:reynolds_inertia}
  \overline{\strut \tau_p \, \widetilde{\bm{a}} \cdot \nabla \widetilde{C} }= \tau_p \left(
    \overline{a}_x \frac{\partial \overline{\strut C}}{\partial x} +
    \overline{a}_z \frac{\partial \overline{\strut C}}{\partial z} + \overline{
      \strut a_x' \frac{\partial c'}{\partial x}} + \overline{\strut  a_z'
      \frac{\partial c'}{\partial z}} \right) 
\end{equation}

\begin{figure*}[htbp]
  \begin{center}
    \includegraphics[scale=1]{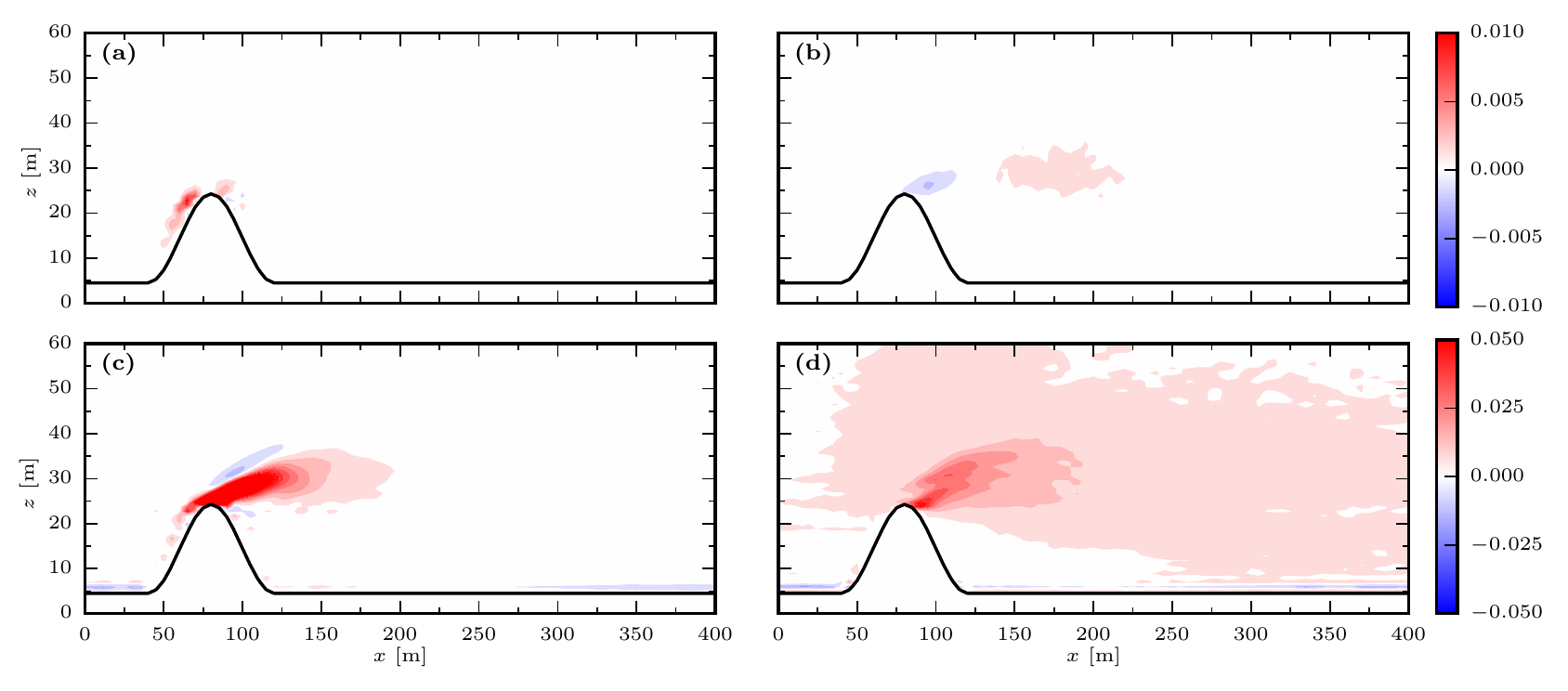}
  \end{center}
  \caption{Mean and turbulent inertia terms in particle equation, averaged in
    time and lateral direction. (a) Mean component of $x$-inertia term, 
    $\frac{L_z}{c_* u_*}\overline{a}_x \frac{\partial \overline{C}}{\partial x}$
    (b) turbulent component of $x$-inertia term, 
    $\frac{L_z}{c_* u_*}\overline{ a_x' \frac{\partial c'}{\partial x}}$'
    (c) Mean component of $z$-inertia term, 
    $\frac{L_z}{c_* u_*}\overline{a}_z \frac{\partial \overline{C}}{\partial x}$
    (d) turbulent component of $z$-inertia term, 
    $\frac{L_z}{c_* u_*}\overline{ a_z' \frac{\partial c'}{\partial z}}$'}
  \label{fig:inertia_terms}
\end{figure*}

Here the $y$ derivatives are neglected due to lateral homogeneity. The average
mean and turbulent $x$- and $z$- inertia terms are displayed in
Fig.~\ref{fig:inertia_terms}, where all terms have been nondimensionalized by
$u_*$, $c_*$, and $L_z$. One can see in panels (a)--(b) that the mean
streamwise inertia leads to the increase concentration of particles on the
windward side of the hill. In contrast, the turbulent component of streamwise
inertia is quite small. The mean $z$-component also contributes to the
increased particle concentration on the windward side of the hill, as well as
to the ``streamer'' of high concentration that extends beyond the crest. Note
here that the contours for the $z$-component are a factor of 5 larger than
those for the $x$-component. The turbulent $z$-inertia component is at its
maximum downwind of the hill at and slightly above hill height, but is smaller
than the mean $z$-component. In contrast to previous studies of preferential
concentration in homogeneous, isotropic turbulence, flows over surface
topography have regions of strong mean accelerations. We find here that for the
simulation considered here, mean inertia dominates over the turbulent
contribution; in fact, the mean component of $z$-inertia is 6-7 times larger
than the turbulent component. This observation is an important detail for
future developments in models of particle deposition in complex terrain, where
the enhancement of deposition can potentially be modeled in terms of mean flow
quantities (e.g. mean concentration and acceleration).

\section{Discussion and Conclusions}\label{sec:conclusions}

A new approach for modeling the transport and deposition of heavy particles
over complex surface topography was developed using the large eddy simulation
technique. An immersed boundary LES code (using the discrete direct forcing
approach) was coupled with an Eulerian finite-volume particle code.  The
advection-diffusion equation for particle mass conservation was discretized
using a Cartesian cut cell method that reshapes finite volume cells intersected
by the zero level set (i.e. the fluid-solid interface). Small cells (with
volume fraction $\alpha < 0.5$) are treated using the conservative mixing model
of \cite{meyer2010conservative} which alleviates the restrictive local CFL
requirements. In order to obtain a divergence-free interpolated velocity field
on finite volume faces intersected by the zero level set, we use a
projection-type method, by first calculating an intermediate interpolated
velocity (with finite divergence) following the approach of
\cite{chamecki2008hybrid}, then solving a Poisson equation to obtain a
divergence-free interpolated velocity.  The proposed numerical method conserves
mass accurately, i.e.~within 0.1\% for simulations of snow deposition over
idealized topography. Accurate mass conservation is essential when simulating
the transport of heavy particles and passive scalars over complex terrain or in
urban environments, motivating our development of a Cartesian cut-cell finite
volume method.

The proposed numerical model was validated by comparing simulation results with
the wind tunnel study of \cite{goossens2006aeolian}, who conducted wind tunnel
experiments of heavy particle deposition over idealized topography.  At fine
($128\times64\times96$) resolution, the LES results were found to be in
reasonably good agreement with observed deposition patterns, although we find
that the results are SGS model-dependent. In particular, the Lagrangian
scale-dependent dynamic model \citep{bouzeid2005scale} is found to overpredict
deposition on the windward side of an obstacle. This is likely related to the
fact that the test filters at scale $2\Delta$ and $4\Delta$ use points inside
the immersed boundary (with nonphysical velocities) to determine a value of the
Smagorinsky coefficient $c_s$ and therefore the SGS scalar diffusivity (in the
present study we use a constant SGS Schmidt number model for the scalar
concentration field). In principle, one could modify the filtering operation so
that only points in the fluid are included in the stencil of the test filters.
Although dynamic SGS models for scalars exist \citep[e.g.][]{stoll2006dynamic},
these too would suffer from the same issue of filtering inside the IBM surface.  

An idealized simulation of snow deposition over a two-dimensional sinusoidal
hill revealed that deposition reached its maximum on the windward side of a
hill, and its minimum on the leeward side, consistent with previous studies of
aeolian processes \citep[e.g.][]{goossens2006aeolian}. The enhanced deposition
on the windward side of the hill can be explained by noting that particle
inertia leads to relative velocities between the particles and fluid, thereby
enhancing concentration on the windward side of the hill. On the leeward side
of the hill, inertia leads to a decreased settling velocity and lower particle
deposition. Reynolds averaging revealed that, for our simulation, the mean
inertia terms in the particle mass conservation equation are larger than their
turbulent counterparts by a factor of 6 or more. 

Our current results and previous studies \citep{lehning2008inhomogeneous,
  mott2010meteorological, mott2010understanding, mott2014orographic} demonstrate
that preferential deposition patterns depend on topographic features, and wind
fields as well as other factors such as atmospheric stratification. We find
from our present results that particle inertia is also significant for
determining the spatial variability of the deposition of heavy particles in
complex terrain.  Our results indicate that Eulerian models of heavy particle
transport in complex terrain should include a model to account for the effects
of particle inertia \citep[e.g.][]{shotorban2007eulerian,
  balachandar2010turbulent} in order to recover physically realistic deposition
patterns. When inertia is excluded, the simulations fail to recover deposition
patterns in quantitative or qualitative agreement with experimental data.


While erosion is not addressed in this work, snow or sediment transport over
topography will also be influence by erosion when the friction velocity $u_*$
exceeds the threshold value $u_{*t}$, thereby initiating saltation
\citep[e.g.][]{bagnold1941physics}. Implementing a model for saltation and
investigating the combined impacts of erosion and deposition will be considered
in future work.

The proposed numerical model conserves mass accurately, is computationally
inexpensive, and is well suited for investigating a variety of problems,
including snow and sediment transport, the dispersal of biogenic particles, and
scalar dispersion in urban environments and complex terrain.

\begin{acknowledgments} MBP gratefully acknowledges support from the National
  Sciences and Engineering Research Council of Canada (NSERC) under the
  Discovery Grants program. The authors thank Elie Bou-Zeid for helpful
  suggestions.  Simulations were run on the Mammouth Parall\`{e}le 2 cluster at
  the Universit\'{e} de Sherbrooke and the Guillimin cluster at McGill
  University, managed by Calcul Qu\`{e}bec and Compute Canada.  The operation
  of these supercomputers is funded by the Canadian Foundation for Innovation
  (CFI), Minist\`{e}re de l'\'{E}conomie, de l'Innovation et des Exportations
  du Qu\`{e}bec (MEIE), RMGA and the Fonds de Recherche du Qu\`{e}bec - Nature
  et Technologies (FRQ-NT). Data from the large eddy simulations used in this
  study are available upon request from the corresponding author (STS,
  salesky@ou.edu).
\end{acknowledgments}

\appendix

\section{Cut Cell Geometry}\label{sec:calc_cut_cell}

\begin{figure}[htbp]
  \begin{center}
    \includegraphics[scale=1.0]{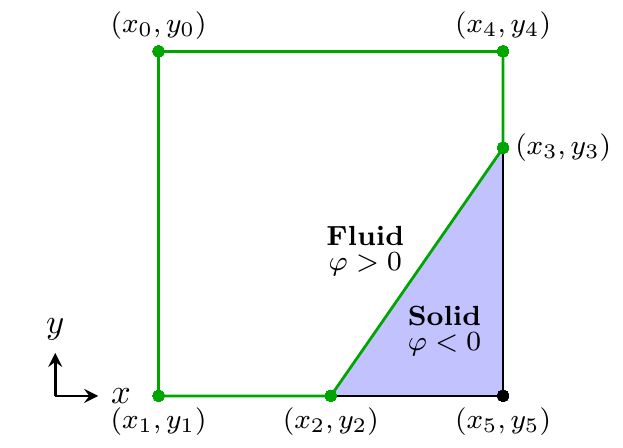}
  \end{center}
  \caption{A schematic diagram of intersection points on the $z$-face of a cut
    cell, used to calculate the area fraction of face in the fluid.}
  \label{fig:cut_face}
\end{figure}

All of the geometric quantities needed for the Cartesian cut cell method (area
fractions of the cut faces, volume fraction of a cell in the fluid, and the
area of a cut face) can be calculated from the level set function $\varphi$.
In order to calculate the area fraction of a cut face, the level set function
is first interpolated to the vertices of that face using trilinear
interpolation (note that $\varphi$ is stored on the $w$ nodes). An example of a
$z$-face intersected by the IBM surface with two of these intersection points,
$(x_2, y_2)$ and $(x_3,y_3)$, where $\varphi=0$, can be found in
Fig.~\ref{fig:cut_face}.

In order to locate the intersection points, $\varphi$ is first interpolated to
the four vertices of a cut face. Faces intersected by the zero level set are
identified where $\varphi$ changes sign between the two adjoining vertices.
For example, the $x$-coordinate of the intersection point on the botton
$y$-face in Fig.~\ref{fig:cut_cell} (i.e. $x_2$) would be located by
\begin{equation}
x_2 = \frac{x_{sw} \varphi_{se} - x_{se} \varphi_{sw}}{\varphi_{se}-\varphi_{sw}}
\end{equation}
where $x_{sw} = x_1 = 0$ and $x_{se} = x_5 = \Delta x$, $\varphi_{sw} =
\varphi(x_1,y_1)$, and $\varphi_{se} = \varphi(x_5,y_5)$.

 Once the intersection points are located, the area of the face in the fluid
 can be calculated using Green's theorem, e.g. for a $z$-face, 
\begin{equation}
    S^z = A^z \Delta x \, \Delta y = \sum_{k=0}^{n} \frac{1}{2} (x_{k+1} + x_k) (y_{k+1} - y_k)
\end{equation}
where the sum is taken over the $n$ vertices of the polygon that surround the part
of the face in the fluid.  For the $z$-face depicted in
Fig.~\ref{fig:cut_face}, one would sum over the vertices $(x_0,y_0),\ldots,
(x_4,y_4)$ to calculate the area of the region outlined in green.

The area of the cut face ($S^{\Gamma}$) can be calculated using de Gua's
theorem (an analogue of the Pythagorean theorem for right tetrahedra): 
\begin{multline}
S^{\Gamma}_{i,j,k} = \bigl[ (S^x_{i+1/2}-S^x_{i-1/2})^2   + 
(S^y_{j+1/2}-S^y_{j-1/2})^2  + \\
(S^z_{k+1/2}-S^z_{k-1/2})^2  \bigr]^{1/2} \text{,}
\end{multline}
and the normal vector on the cut face can be calculated via:
\begin{equation}
  \bm{n^{\Gamma}_{i,j,k}} =  \frac{1}{S^{\Gamma}} \left(\begin{matrix}
  S^{x}_{i+1/2}-S^{x}_{i-1/2} \\
  S^{y}_{j+1/2}-S^{y}_{j-1/2} \\
  S^{z}_{k+1/2}-S^{z}_{k-1/2}
\end{matrix}\right)
\end{equation}
Once the face areas have been calculated, the volume fraction of a cut cell can be calculated using the 
divergence theorem:
\begin{equation}
  \alpha_{i,j,k} = \frac{1}{\Delta x \ \Delta y \, \Delta z} \sum_{\ell}
  \frac{1}{3} \left(\bm{x}_\ell \cdot \bm{n}_\ell S_\ell + \bm{x}_{\Gamma} \cdot
    \bm{n}_{\Gamma} S_{\Gamma} \right)
\end{equation}
where $\bm{x_\ell}$, $\bm{n_\ell}$, and $S_\ell$ are respectively a point on
the $\ell$th face, the outward facing normal vector on that face, and the face
area; $\bm{x_\Gamma}$, $\bm{n}_\Gamma$, and $S_\Gamma$ represent the
corresponding quantities on the cut face (e.g. the sum would be taken over
regular faces $\ell=1$, \ldots, $6$ and the cut face for the finite volume cell
depicted in Fig.~\ref{fig:cut_cell}).  For the case of a fixed surface height
as in the present work (i.e. where the level set is not advected), all
geometric quantities are calculated once at the beginning of the computation
and stored.

\section{Conservative Mixing Model}\label{sec:conserv_mix}

In the finite volume Cartesian cut cell code, we employ the conservative mixing
model of \citet{meyer2010conservative} to address the so-called ``small cell
problem'' to circumvent the need to decrease the timestep in cut cells with
small volume fractions in order to satisfy the local CFL condition.  The mixing
procedure is performed each timestep after AB2 time advancement, and is
summarized here briefly.  The concentration exchanged between a small cell
$(i,j,k)$ and a target cell in the positive $x$-direction $(i+1,j,k)$, is given
as:
\begin{equation}
X^x = \frac{\beta_{i,j,k}^x  \mathcal{V}_{i,j,k} \mathcal{V}_{i+1,j,k} }{\beta_{i,j,k}^x \mathcal{V}_{i,j,k} +
  \mathcal{V}_{i+1,j,k}}  \left[
  \widetilde{C}_{i+1,j,k} - \widetilde{C}_{i,j,k} \right]
\end{equation}
where $\mathcal{V}_{i,j,k} = \Delta x \,\Delta y\, \Delta z\, \alpha_{i,j,k}$
is the volume of cell $(i,j,k)$ and $\beta^x_{i,j,k}$ is a mixing fraction,
defined below.  Similar terms are calculated for the other neighboring cells; a
small cell will mix concentration with seven neighboring cells in 3D.  The
mixing fractions are given as
\begin{equation}
  \begin{aligned}
    \beta^x_{i,j,k} = | n^{\Gamma}_x n^{\Gamma}_x | (\alpha_{target}^x)^\mu \\
    \beta^y_{i,j,k} = | n^{\Gamma}_y n^{\Gamma}_y | (\alpha_{target}^y)^\mu \\
    \beta^z_{i,j,k} = | n^{\Gamma}_z n^{\Gamma}_z | (\alpha_{target}^z)^\mu \\
    \beta^{xy}_{i,j,k} = | n^{\Gamma}_x n^{\Gamma}_y | (\alpha_{target}^{xy})^\mu \\
    \beta^{xz}_{i,j,k} = | n^{\Gamma}_x n^{\Gamma}_z | (\alpha_{target}^{xz})^\mu \\
    \beta^{yz}_{i,j,k} = | n^{\Gamma}_y n^{\Gamma}_z | (\alpha_{target}^{yz})^\mu \\
    \beta^{xyz}_{i,j,k} = |n^{\Gamma}_x n^{\Gamma}_y n^{\Gamma}_z |^{2/3} (\alpha_{target}^{xyz})^\mu 
  \end{aligned}
\end{equation}
where $\bm{n^\Gamma} = (n^\Gamma_x, n^\Gamma_y,n^\Gamma_z)$ is the normal
vector on the cut face, $\alpha_{target}^x$ is  the volume fraction of the
``target cell'' for mixing and is determined by the direction of the normal
vector, i.e.
\begin{equation}
 \alpha_{target}^x  = 
 \begin{cases}
   \alpha_{i+1,j,k}, & n^\Gamma_x > 1 \\
   \alpha_{i-1,j,k}, & n^\Gamma_x < 1 
 \end{cases} \text{,}
\end{equation}
and $\mu \geq 1$ is an integer to give a larger weight to cells with large
volume fractions for the sake of numerical stability.
\citet{meyer2010conservative} used a value of $\mu = 5$; we find little
sensitivity to the value of $\mu$ employed.  Note that the mixing fractions
normalize to $1$, i.e.
\begin{equation}
  \beta^x_{i,j,k} + \beta^y_{i,j,k} +\beta^z_{i,j,k} +\beta^{xy}_{i,j,k}
  +\beta^{xz}_{i,j,k} +\beta^{yz}_{i,j,k} +\beta^{xyz}_{i,j,k} = 1
\end{equation}
 After calculating the mixing fractions and exchanging concentrations, the
 concentration in the small cells and target cells are updated, e.g.
  in a target cell via
 \begin{equation}
   \widetilde{C}_{i+1,j,k} = \widetilde{C}^\star_{i+1,j,k} - \frac{1}{\mathcal{V}_{i+1,j,k}} X^x
 \end{equation}
and in the small cell by
 \begin{equation}
   \widetilde{C}_{i,j,k} = \widetilde{C}^\star_{i,j,k} + \frac{1}{\mathcal{V}_{i,j,k}} \left[
     X^x + X^y + X^z + X^{xy} + X^{xz} + X^{yz} + X^{xyz}
   \right]
 \end{equation}
 where $\widetilde{C}^\star_{i,j,k}$ is the value of concentration before conservative mixing.


\bibliographystyle{agufull08.bst}      
\bibliography{refs.bib}   

\end{article}

\end{document}